\documentstyle[12pt,aasms4,astrobib]{article}
\slugcomment{Submitted to the Astrophysical Journal, 16 May 1997}

\lefthead{Isenberg, Lamb, and Wang} \righthead{Cyclotron Line
Formation in a Radiation-driven Outflow}

\begin{document} 

\title{Cyclotron Line Formation in a Radiation-driven Outflow}

\author{Michael Isenberg and D. Q. Lamb} \affil{Department of Astronomy
and Astrophysics, University of Chicago, 5640 South Ellis Avenue,
Chicago, IL 60637}

\author{John C. L. Wang} \affil{Department of Astronomy, University of
Maryland, College Park, MD 20742-2421}

\begin{abstract} 
We calculate the properties of gamma-ray burst spectral lines formed by
resonant cyclotron scattering in a radiation-driven outflow.  Most
previous models of line formation in gamma-ray bursts are appropriate
at the polar cap of a neutron star located no further than several
hundred parsecs away.  However, the BATSE brightness and sky
distributions indicate that, if bursters are galactic, they are
located in a corona at distances greater than 100 kpc.  At these
distances the burst luminosity exceeds the Eddington luminosity and
the plasma in a polar line-forming region is ejected along the field
lines.  The variation of the magnetic field strength and plasma
velocity with altitude in such an outflow would seem to prevent the
formation of narrow features like those observed by {\it Ginga} and
other instruments.  However, this is not the case because the majority
of scatters occur close to the stellar surface, at altitudes below $z
\approx r_h$, where $r_h \sim 10^{5}$ cm is the size of the photon
source.  Consequently, the interpretation of the observed features as
cyclotron lines does not rule out burst sources in a galactic corona.

The outflow model predicts both absorption-like and single-peaked
emission-like features.  The latter do not occur in models with static
line-forming regions, but have been observed in gamma-ray bursts by
the Konus and BATSE instruments.
\end{abstract}

%%%%%%%%%%%%%%%%%%%%%%%%%%%%%%%%%%%%%%%%%%%%%%%%%%%%%%%%%%%%%%%%%%%
%                                                                 %
%     INTRODUCTION                                                %
%                                                                 %
%%%%%%%%%%%%%%%%%%%%%%%%%%%%%%%%%%%%%%%%%%%%%%%%%%%%%%%%%%%%%%%%%%%

\section{Introduction}

The source of gamma-ray bursts is one of the most hotly debated
questions in astrophysics (\citeNP{lamb95}; \citeNP{paczynski95}).  These
short, intense bursts of high energy radiation were first observed by
the Vela 3 and 4 satellites on July 2, 1967 (\citeNP{kso73};
\citeNP{bk96}).  Today, the main question continues to be the distance
to the sources -- are they galactic or cosmological?  Interest in this
mystery has intensified in recent months due to the hard gamma-ray
repeater candidate of October 27-29 1996 and the spectacular follow-up
observations to gamma-ray burst GB970228.  The October 1996 observations are a
sequence of four events over a two day period with positions in the
sky that are consistent with a single source (\citeNP{meegan96}).  They
are the best evidence to date that burst sources can repeat, and that
therefore bursting does not necessarily lead to the destruction of the
source.  X-ray and optical observations of the GB970228 field offer
the best candidates to date for burst counterparts in other
wavelengths.  These observations reveal a transient x-ray source
(\citeNP{costa97}), a transient optical point source (\citeNP{groot97a}),
and an extended optical source (\citeNP{groot97b}; Metzger et al.,
1997a,b\nocite{metzger97a}\nocite{metzger97b}; \citeNP{sahu97}) whose
positions in the sky are consistent with that of the burst and with
each other.

\medskip
One of the strongest pieces of evidence in favor of the galactic point
of view is the observation of statistically significant absorption-
and emission-like features in the spectra of some bursts, and the
interpretation of these features as cyclotron lines (Mazets et
al. 1981, 1982\nocite{mazets81}\nocite{mazets82}; \citeNP{lamb82};
\citeNP{bs82}; \citeNP{mazets83}; \citeNP{hueter84};
\citeNP{hueter87}; \citeNP{murakami88}; \citeNP{fenimore88};
\citeNP{hp89}; \citeNP{wang89}; \citeNP{am89}; \citeNP{lamb89};
\citeNP{graziani92}; \citeNP{yoshida92}; \citeNP{ne92};
\citeNP{barat93}; \citeNP{nishimura94}; \citeNP{mazets96};
\citeNP{briggs96a}; \citeNP{briggs96b}).  Cyclotron lines are formed
by the resonant scattering of photons by electrons in a magnetic
field, $B$.  The motion of free electrons perpendicular to the field
is quantized into energy levels (\citeNP{landau30}; \citeNP{tbz66}),

\begin{equation}
E_e = \sqrt{1+2nb}\ m_e c^2 
\label{defEn}
\end{equation}

\noindent where $n$ is an integer $\geq 0$, $b \equiv B/B_c$, $B_c=4.414 \times
10^{13}$ gauss is the critical field where the cyclotron energy $E_B
\equiv b\ m_e c^2  = 11.6\ B_{12}$ is equal to the electron rest energy,
and $B_{12} \equiv B/10^{12}$ gauss.  When $b \ll 1$, the energy
level spacing is approximately harmonic, i.e. $E_e \approx m_e c^2 + n E_B$.
Photons which
possess the right energy to excite an electron to a higher Landau
level can scatter resonantly into or out of the line of sight,
producing spectral features.

\medskip
Although the observation of spectral features in gamma-ray bursts and
their interpretation continue to be controversial, the existence of
Landau levels and the formation of cyclotron lines in the spectra of
{\it persistent} astrophysical sources with strong magnetic fields are
well established experimentally.  The quantization of electron
energies in metals is responsible for the de Haas-van Alphen effect --
the oscillation of magnetization with magnetic field strength
(\citeNP{dv30}).  Cyclotron lines appear in the x-ray
spectra of about a dozen accretion-powered pulsars (\citeNP{mihara95};
\citeNP{mm92}).  These sources include Her X-1 (\citeNP{trumper78}), 4U
0115+63 (\citeNP{wheaton79}), 4U 1538-52 (\citeNP{clark90}), and A0535+26
(\citeNP{grove95}).  Visvanathan and Wickramasinghe
(1979\nocite{visvan79}) made the first detection of cyclotron lines
from a white dwarf in infrared observations of the AM Her type binary
VV Pupis.  Since then cyclotron lines have been detected in about a
dozen AM Her binaries, including AM Her itself (for a review see
\citeNP{chanmugam92}).  Comparing these observations with theoretical
models of line-formation reveals much about the physical conditions in
the line forming region, including the magnetic field strength, the
plasma density, and the temperature.

\medskip
The lines reported in gamma-ray bursts are at
energies $\approx 20-60$ keV, corresponding to magnetic fields $B_{12} \approx
2-6$, consistent with a neutron star source.  The {\it Ginga}
satellite's observation of {\it harmonically spaced} features in three
bursts strongly supports the cyclotron line interpretation.  The
features persist for as long as 9 seconds (\citeNP{graziani92}).  If the
cyclotron line interpretation is correct, then the magnetic field must
remain steady at least as long as the line is observed.  The
requirement of a steady field rules out some cosmological models, such
as coalescing neutron stars, which require the catastrophic
annihilation of the source.

\medskip
Establishing that the observed features are indeed cyclotron lines
requires comparing the observations with theoretical models that are
appropriate for the physical conditions expected at the source.  Most
previous calculations of line formation in gamma-ray bursts, such as
those of Wang et al. (1989\nocite{wang89}), model the line-forming
region as a static slab of plasma, threaded with a uniform magnetic
field oriented perpendicular to the surface of the slab.  Such a model
is appropriate, for example, at the magnetic polar cap of a neutron
star with a dipole field, located less than several hundred parsecs
from the earth.  Lamb, Wang, and Wasserman (1990\nocite{lww90}) show
that for larger distances the static polar cap model is not valid; the
burst luminosity exceeds the Eddington luminosity and the radiation
force creates a relativistic plasma outflow along the field lines.

\medskip
However, in order to explain the burst brightness and sky
distributions observed by BATSE, it has been suggested that, if the
bursters are galactic, they are located in a galactic corona at
distances of 100-400 kpc (for a review, see \citeNP{lamb95}).  Two
models have been suggested that are appropriate for these distances:
an equatorial model and an outflow model.

\medskip
In the equatorial model, the line-forming region is located at the
magnetic equator, where field lines parallel to the stellar surface
confine the plasma magnetically (see, e.g. \citeNP{lamb82};
\citeNP{katz82}; Zheleznyakov and Serber 1994,
1995\nocite{zs94}\nocite{zs95}).  Freeman et al.
(1996)\nocite{freeman96} fit both this model and a static polar cap one to
the two observed spectra corresponding to the time intervals S1 and S2
during which lines are observed in GB870303.  A joint fit to the two
intervals, using models with a common magnetic field and column depth,
but not a common viewing angle, marginally favors the equatorial model
over the static polar cap one.

\medskip
In the outflow model the radiation force on the plasma at the polar
cap ejects the plasma relativistically along the field lines.  Table
1 compares the existing calculations of the properties of
cyclotron lines formed in such an outflow.  The variation of magnetic
field strength and plasma velocity with altitude would seem to prevent
the formation of narrow scattering lines.  However, Miller {\it et
al.} (1991\nocite{miller91}, 1992\nocite{miller92}) show that narrow
lines can be formed in an outflow at the second and third harmonics,
which they approximate as due to cyclotron {\it absorption}.
Chernenko and Mitrofanov (1993\nocite{cm93}) calculate the properties
of the first harmonic line, also approximating it as due to
absorption, and find that the formation of a narrow line is possible.
However, the absorption approximation is not valid for the first
harmonic; multiple photon scatters must be taken into account.  The
first calculation to do so was Isenberg, Lamb, and Wang
(1996\nocite{ilw96}) using a Monte Carlo radiative
transfer code.  This calculation includes multiple scattering at the
first three harmonics in a plasma with electron temperature $k_B
T=0.25\ E_B$.  It shows lines with equivalent widths comparable to
those observed by {\it Ginga}.

\medskip
The emerging photon spectrum is very sensitive to the velocity profile
$\beta_F(z)$ -- the bulk flow velocity as a function of the altitude
$z$ above the stellar surface.  An electron with velocity $\beta$
along the field is overtaken by photons with $\mu > \beta$, where
$\mu=\cos \theta$, and $\theta$ is the angle between the photon's
direction of travel and the magnetic field.  In the electron's rest
frame, these photons have orientations $\mu^r>0$; scattering with them
accelerates the electron.  Similarly, photons with $\mu < \beta$ have
rest frame orientations $\mu^r < 0$ and decelerate the electrons.
Consequently, the three previous calculations assume that the flow
velocity at a given altitude is approximately the mean value of the
component of the photon velocity along the field; that is

\begin{equation}
\beta_F(z) \approx {\int{\mu\ Q(\mu,z) d\mu} \over {\int{Q(\mu,z)}
d\mu}}\ ,
\label{bfapprox1}
\end{equation}

\noindent
where $Q(\mu,z)$ is the angular distribution of photons at altitude
$z$.  This profile is a good approximation to the one Tsygan
(1981)\nocite{tsygan81} and Mitrofanov and Tsygan (1982; hereafter
MT)\nocite{mt82} derive from the radiation force $F_{rad}(z)$.  Their
derivation includes scattering at the first harmonic only, with $k_B
T=0$.  It is based on the {\it unscattered} photon distribution; i.e.,
the calculation of the force at $z$ does not take into account the
reduction in photon flux near the line center or the angular
redistribution of photons due to scattering at lower altitudes.  This
approximation is valid at small optical depths, but not at the depths
required to reproduce the observed cyclotron lines.  Miller {\it et
al.} (1991, 1992) and Isenberg, Lamb, and Wang (1996) also include
processes in their radiation transfer calculations, such as higher
harmonic scattering and finite temperature, which are not included in
MT's calculation of $F_{rad}$.  Thus, in the earlier models, the
hydrodynamics is not consistent with the radiation transfer.
	
\medskip
In the current work, we present the first {\it self-consistent}
calculations of the properties of cyclotron lines formed in an
outflow.  In addition to the physical processes included in Isenberg,
Lamb, and Wang (1996), we add
continuum scattering, finite natural line width, and photon
polarization.  We describe the geometry and the physics of the line
forming region in sections \ref{lfrgeo} and \ref{lfrphys}
respectively.  The electron velocity profiles are discussed in section
\ref{bprof}.  We present the Monte Carlo spectra in section
\ref{mcspec} and discuss their implications for gamma-ray bursts in
section \ref{discuss}.
  
%%%%%%%%%%%%%%%%%%%%%%%%%%%%%%%%%%%%%%%%%%%%%%%%%%%%%%%%%%%%%%%%%%%
%                                                                 %
%     Spatial Geometry of the Line-Forming Region                 %
%                                                                 %
%%%%%%%%%%%%%%%%%%%%%%%%%%%%%%%%%%%%%%%%%%%%%%%%%%%%%%%%%%%%%%%%%%%

\section{Spatial Geometry of the Line-Forming Region}
\label{lfrgeo}
We adopt a cylindrical line-forming region, as shown in figure \ref{lfr}.
Photons are injected at the base of the cylinder, which could
correspond to a
hot spot with radius $r_h$ located on the surface of a neutron star.
The angular size of the hot spot is represented by

\begin{equation}
\mu_o(z) = \cos \theta_o(z) = {\tilde{z} \over (1 +
\tilde{z}^2)^{1/2}}\ ,
\label{defmu0}
\end{equation}

\noindent where $z$ is the altitude above the surface and
the tilde indicates quantities measured in units of $r_h$.

\medskip
We assume the star has a dipole magnetic field with strength $B_o$,
where the subscript $o$ indicates quantities measured at the stellar
surface.  We also assume that the
hot spot is located on the stellar surface, at the magnetic pole.  The
field strength decreases with $z$ according to 

\begin{equation}
B=B_o\ \left ({1 \over 1  + z/R_*} \right )^3\ ,
\label{bdip}
\end{equation}

\noindent
where $R_*$ is the stellar radius.
The field lines that are rooted in the hot spot form a tube with a
circular cross section with radius $\tilde{r}(z)$.
Conservation of magnetic flux requires that $\tilde{r}(z)$ increase
with altitude according to

\begin{equation}
\tilde{r}(z)=\sqrt{B_o \over B(z)} = \left ( 1+{z \over R_*}\right
)^{3/2}\ , 
\label{flare}
\end{equation}

\noindent i.e. the field lines flare outwards.  For simplicity, we
assume the field lines are parallel to the cylinder axis at all $z$; we
ignore the flaring in our Monte Carlo simulations, except when
calculating the plasma density.  The characteristic distance for
changes in $\tilde{r}(z)$ is $\sim (d\tilde{r}/dz)^{-1} \sim R_*$.
But, as we shall show, few scatters occur above $z \sim r_h \ll R_*$.
Therefore, the effect of this assumption on the emerging photon
spectrum is negligible, since most of the scattering takes place close
to the surface where $\tilde{r}(z)$ is approximately constant.

\medskip
The line-forming region contains a plasma in which the number density
of scatterers (electrons and positrons) is $n_e(z)$.  The plasma can
consist of ionized electron-proton pairs that have been swept up from
the stellar surface or electron-positron pairs that have been created
by the processes $\gamma \rightarrow e^- e^+$ and $\gamma\gamma
\rightarrow e^- e^+$.  Our Monte Carlo simulations assume the
number of scatterers is conserved; i.e. the simulations apply to
line-forming regions where electron-proton and electron-positron
plasmas originate at the surface.  The effects of pair creation and
annihilation above the surface is beyond the scope of our current
calculations (although, see section \ref{discuss} for a general
discussion; see also \citeNP{vd91}; \citeNP{cl92}).  Because the 
electron temperature $k_B T_e \ll E_B$, the electrons are restricted to
move along the field lines (see section \ref{sPdistf}).  As the
photons travel through the plasma they can scatter with the electrons
(and positrons), accelerating them.  We denote
the velocity along the field lines of an individual electron (in units
of $c$) by $\beta$ 
and the bulk flow velocity by $\beta_F(z)$.

\medskip
Photons which cross the lateral surface of the cylinder escape the
line-forming region and reach the observer.  In the absence of
scattering, a photon injected at the origin escapes when
$\mu_o(z)=\mu$.  The resulting angular distribution of photons is

\begin{equation}
Q(\mu,z)=Q(\mu)\ \eta \bigg [ \mu-\mu_o(z) \bigg ]\ ,
\label{angledist}
\end{equation}

\noindent where $Q(\mu)$ is the angular distribution at the stellar
surface and $\eta$ is the unit step function.  For simplicity, we use
semi-isotropic injection ($Q(\mu)=1$) throughout the present work.
Note however that some models of spectrum formation in gamma-ray
bursts suggest that the photons are beamed along the field lines (see
e.g. \citeNP{schmidt78}; \citeNP{bh93}; \citeNP{hb94}; \citeNP{harding94};
Section \ref{discuss} below).  This possibility should be considered
in future calculations of line-formation in an outflow.  Also note that the
distribution given by eq. (\ref{angledist}) for photons injected at
the origin of the hot spot is also the angular distribution along the
cylinder axis when photons are injected uniformly across the disk.
Substituting this distribution into eq. (\ref{bfapprox1}) yields the
approximate velocity profile

\begin{equation}
\beta_F(z) \approx {1 \over 2}\ \bigg [1+\mu_o(z) \bigg ]
\label{bfapprox2}
\end{equation}

\medskip The few photons
which reach $\tilde{z} > 30$ are also permitted to
escape without further scatters; for $r_h \sim 0.1 R_*$, the plasma is
extremely tenuous above this altitude and the cyclotron energy less
than the energy threshold of most gamma-ray burst detectors.  We assume
that photons which scatter downwards and return to the surface
thermalize, i.e. they are absorbed by nonresonant inverse magnetic
bremsstrahlung.

%%%%%%%%%%%%%%%%%%%%%%%%%%%%%%%%%%%%%%%%%%%%%%%%%%%%%%%%%%%%%%%%%%%
%                                                                 %
%     Physics of the Line-Forming Region                          %
%                                                                 %
%%%%%%%%%%%%%%%%%%%%%%%%%%%%%%%%%%%%%%%%%%%%%%%%%%%%%%%%%%%%%%%%%%%

\section{Physics of the Line-Forming Region}
\label{lfrphys}

In this section we describe the basic physics that govern the
properties of cyclotron lines formed in outflows: the characteristic
time scales (section \ref{stimes}), the one dimensional electron
momenta distribution $f(p)$ (section \ref{sPdistf}), the plasma
density profile $n_e(z)$ (section \ref{spdens}), the scattering
kinematics (section \ref{skine}), the scattering profile
$\phi(E,\mu,z)$ (section \ref{sprofiles}), and the gravitational red
shift (section \ref{srshift}).  We discuss the electron velocity
profile $\beta_F(z)$ separately, in section \ref{bprof}.

%%%%%%%%%%%%%%%%%%%%%%%%%%%%%%%%%%%%%%%%%%%%%%%%%%%%%%%%%%%%%%%%%%%
%                                                                 %
%     Characteristic time scales                                  %
%                                                                 %
%%%%%%%%%%%%%%%%%%%%%%%%%%%%%%%%%%%%%%%%%%%%%%%%%%%%%%%%%%%%%%%%%%%

\subsection{Characteristic time scales}
\label{stimes}

Five time scales characterize the physics of the line-forming region:
the radiative lifetime $t_{rad}$, the electron-photon relaxation time
scale $t_{e\gamma}$, the electron-proton and proton-proton relaxation
time scales $t_{ep}$ and $t_{pp}$, and the particle escape time
$t_{esc}$.  For simplicity, we estimate the first four time scales for
a {\it static} [$\gamma_F \equiv (1-\beta_F^2)^{-1/2} = 1$]
line-forming region.  This approximation is suitable near the stellar
surface, where $\gamma_F < 1.2$.  All these time scales are short
compared to the burst duration $\sim$ 1-10 s.

\medskip
The radiative lifetime of an electron in the $n$th Landau level is
just 

\begin{equation}
t_{rad}={\hbar \over \Gamma_n}=3\times 10^{-16}\ {\rm s}\ n^{-1}
B_{12}^{-2}\ ,
\label{trad}
\end{equation}

\noindent where

\begin{equation} 
\Gamma_n = {4 n \alpha E_B^2 \over 3 m_e c^2}=2.6 \times 10^{-3} \ {\rm keV}\ nB_{12}^2
\end{equation}

\noindent
is the natural line width for the $n$th harmonic and $\alpha$ is the
fine structure constant.  

\medskip
We determine the time scale for the electrons to come to equilibrium with the
photons by calculating the Fokker-Planck coefficient

\begin{equation}
\left \langle {\delta p^2 \over \delta t} \right \rangle_{e\gamma}
=\int{dE \int{ {d\Omega \over 2\pi}\ n_x(E,\Omega) \int{d\Omega_s {d\sigma
\over d\Omega_s} c\ \delta p^2}}}\ ,
\label{ddpdt}
\end{equation}

\noindent where 

\begin{equation}
\delta p \approx (\mu-\mu_s)\ E
\label{delp}
\end{equation}

\noindent
is the change in momentum during a scattering, the subscript $s$
denotes parameters of the scattered photon, and

\begin{equation}
n_x(E,\Omega,z)\ dE {d\Omega \over 2\pi} =Q(\mu,z)\ 
n_x(E,z)\ dE { d\Omega \over 2\pi}
\label{ngammaEO}
\end{equation}

\noindent
 is the density of x-ray photons at altitude
$z$  with 
energy between $E$ and $E+dE$ and direction between $\Omega$ and
$\Omega+d\Omega$.
Assuming first harmonic scattering dominates the radiation
transfer and taking the limit of zero natural line width, the
polarization averaged scattering cross section is, to
lowest order in $\beta$ and $b$, approximately

\begin{equation}
{d\sigma \over d\Omega_s} \approx {9 \over 32} \sigma_T {m_e c^2 \over
\alpha}\ \delta (E-E_B)\ {1+\mu^2 \over 2}\ {1+\mu_s^2 \over 2}
\label{xsecn1st}
\end{equation}

\noindent (see Section \ref{sprofiles}), where $\sigma_T\equiv
(8\pi/3)[e^2/(m_e c^2)]^2=6.65 \times 10^{-25}\ {\rm cm^2}$ is the
Thomson cross section.  Substituting eqs.
(\ref{delp}) -- (\ref{xsecn1st}) into eq. (\ref{ddpdt}) yields 

\begin{equation}
\left \langle {\delta p^2 \over \delta t} \right \rangle_{e\gamma}
={13 \pi \over 160}\ \sigma_T {m_e c^3 \over \alpha} n_x(E_B)\ .
\end{equation}

\noindent The equilibration time scale is

\begin{equation}
t_{e\gamma} \approx {\delta p^2 \over \left \langle \delta p^2 /
\delta t \right \rangle_{e\gamma}}\sim {(E_B/c)^2 \over \left \langle \delta p^2 /
\delta t \right \rangle_{e\gamma}}\ = {160\ \alpha \over 13 \pi\ 
\ \sigma_T\ m_e c^3\ n_x(E_B)}\ .
\label{tegdef}
\end{equation}

\noindent
If the electron column depth is sufficiently low so
that photons scatter at most one time before escaping the line-forming
region (the single scattering approximation), then at all $z$
the photon spectrum is approximately the same as the spectrum 
initially injected at the hot spot.  For a power law spectrum with
spectral index $s$,

\begin{eqnarray}
n_x(E,z)=n_x(E)& = & {L_x \over \pi r_h^2 c\ \xi(s)}\ E^{-s}
\nonumber \\
& = & 6.6
\times 10^{27}\ {\rm keV\ cm^{-3}}\ {L_{x,40} \over r_{h,5}^2\ \xi(s)}
\left ( {E \over 1\ {\rm keV}} \right )^{-s}\ , 
\label{ngammaE}
\end{eqnarray}

\noindent
where $r_{h,5} \equiv r_h/10^5$ cm, $L_{x,40} \equiv L_x/10^{40}\ {\rm
erg\ s^{-1}}$ 
is the luminosity of x-ray photons with $E_1<E<E_2$, and

\begin{equation}
\xi(s) \equiv \int_{E_1}^{E_2}{E^{1-s} dE}\ .
\label{xis}
\end{equation}

\noindent
Throughout this paper we use $E_1$=1 keV and $E_2$=1000
keV.  Substituting eqs. (\ref{ngammaE}) and (\ref{xis}) into
eq. (\ref{tegdef}) and choosing $s=1$ [$\xi(s)\approx 1000$ keV],

\begin{equation}
t_{e\gamma} \sim 5 \times 10^{-15}\ {\rm s}\ L_{x,40}^{-1}\ r_{h,5}^2
\ B_{12}\ .
\label{teg}
\end{equation}

\medskip
By similar arguments, the relaxation time for electron-proton
scattering is

\begin{equation}
t_{ep} \sim 3 \times 10^{-6}\ {\rm s}\ n_{p,17}^{-1}\ T_k^{3/2}
\left ( \ln \Lambda_{ep} \over 20 \right )^{-1}\ ,
\end{equation}

\noindent where $T_k\equiv k_B T_e / 1$ keV, the Coulomb
logarithm is $\ln \Lambda_{ep} \approx 16.2 + 2\ \ln (T_k B_{12}^{-1})$
(\citeNP{langer81}), and $n_{p,17} \equiv n_p/10^{17} \ {\rm cm^{-3}}$
is the proton 
number density ($=n_{e,17} \equiv n_e/10^{17} \ {\rm cm^{-3}}$ in a pure
hydrogen plasma).  
We choose a characteristic density, $10^{17}\ {\rm cm^{-3}}$, 
where the line-forming region is optically thick in the
line core but optically thin in the wings (see section \ref{sprofiles}
below, esp. eqs. [41] and [47] and associated discussion).  
The relaxation time for proton-proton scattering is

\begin{equation}
t_{pp} \sim 2 \times 10^{-7}\ {\rm s}\ n_{p,17}^{-1}\ T_k^{3/2}
\left ( \ln \Lambda_{ep} \over 20 \right )^{-1}\ ,
\end{equation}

\noindent where $\ln \Lambda_{pp} \approx
15.9 + 0.5\ \ln(T_k\ n_{p,17}^{-1})$ (\citeNP{spitzer62}).  Since the
electrons are confined to motion along 
the field lines, they merely exchange momentum when they
scatter with each other, with no change to the overall momenta
distribution.  

\medskip
Finally, the time required for an electron to escape the region near
the surface, where most scatters take place, is

\begin{equation}
t_{esc} \sim {r_h \over c} = 3 \times 10^{-6}\ {\rm s}\ r_{h,5}\ .
\end{equation}

\medskip
Comparing the time scales reveals a great deal about what physical
processes are important in the line-forming region.  Since
$t_{e\gamma} \ll t_{ep,pp}$, radiation processes 
dominate collisional processes.  The reason for this is clear:
the density of electrons is small compared to the density
of resonant photons.  While the
electron number density in the outflow model is $10^{17}-10^{19} \
{\rm cm^{-3}}$, the density of resonant photons is $\approx
n_x(E_B)\ E_d^1 = 4 \times 
10^{23}\ {\rm cm^{-3}}\ L_{x,40}\ T_k^{1/2}\ r_{h,5}^{-2}$, where

\begin{equation}
E_d^n \equiv n E_B\ \sqrt{2 T_e \over m_e c^2} = 0.73\ n B_{12}  T_k^{1/2}
\end{equation}

\noindent 
is the Doppler width associated with the $n^{th}$ harmonic.  Since
$t_{ep,pp}$ may be $\gtrsim t_{esc}$, the particles may not spend enough
time in the line-forming region to reach equilibrium by
particle-particle collisions.  However, $t_{e\gamma} \ll
t_{esc}$, so there is ample time for the electron momenta distribution to
reach equilibrium by electron-photon scatterings, as we discuss in more
detail in Section \ref{sPdistf}.

\medskip
Since particle-particle collisions are rare, radiative processes
determine the population of excited Landau states.  As
eqs. (\ref{trad}) and (\ref{teg}) show, for most situations of
interest for line formation in gamma-ray bursts the radiative decay
time of the excited states is shorter than the mean time between
electron-photon scatters.  Consequently the excited states are not
radiatively populated.  Thus, we assume throughout this work that an
electron is in the Landau ground state ($n=0$) before and after a
scattering.  Note, however, that this assumption breaks down for low
field strengths and high luminosities where $B_{12}^3\ L_{x,40}^{-1}
\lesssim 0.1$.

\medskip
The low rate of particle-particle collisions suggests that pure
absorption (i.e. nonresonant inverse magnetic bremsstrahlung) is rare.  A
more careful analysis confirms this.  For $E_B \gg k_B T_e$, the
absorption probability {\it per scattering} is the ratio of absorption
to scattering cross sections 

\begin{eqnarray}
P_a=\left [ {\sigma_a \over \sigma_s} \right ]_{B \neq 0} \approx
\left [ {\sigma_a \over \sigma_T} \right ]_{B=0} & \approx & 2 \pi \alpha\
\sqrt{m_e c^2 \over E_B}\ {\hbar^3 c^3 n_p \over E_B^3} 
\nonumber \\
& \approx & 1.5 \times 10^{-10}\ n_{p,17}\ B_{12}^{-7/2}
\end{eqnarray}

\noindent (\citeNP{nelson93}).  If the line wings are optically thin
(see Section \ref{sprofiles}), a
photon in the line core can escape the line-forming region by
scattering to the wings.  A core-wing transition occurs in
$\sim 1/a$ scatters, where

\begin{equation}
a \equiv {\Gamma_1 \over 2 E_d^1}=1.8 \times 10^{-3}\ B_{12} T_k^{-1/2}
\label{natline}
\end{equation} 

\noindent
is the dimensionless natural line width (\citeNP{ws80}).  The total
absorption probability for a resonant photon is, therefore, $a^{-1}
P_a \sim 10^{-7}\ n_{p,17}\ B_{12}^{-9/2}\ T_k^{1/2}$.  Since this
probability is $\ll 1$ throughout the current work, we do not include
pure absorption in our calculations.

%%%%%%%%%%%%%%%%%%%%%%%%%%%%%%%%%%%%%%%%%%%%%%%%%%%%%%%%%%%%%%%%%%%
%                                                                 %
%     Electron Momenta Distribution                               %
%                                                                 %
%%%%%%%%%%%%%%%%%%%%%%%%%%%%%%%%%%%%%%%%%%%%%%%%%%%%%%%%%%%%%%%%%%%

\subsection{Electron Momenta Distribution} \label{sPdistf}

The time scales that govern the relaxation of electrons in a dynamic,
galactic corona model are dramatically different from those of a
static, galactic disk model where the photon densities are much lower
and the electrons do not escape the line-forming region.
Consequently, determining the distribution of electron momenta along
the field lines in an outflow requires some care.  Two questions must
be answered: what is the equilibrium distribution and are the
electrons in equilibrium --- i.e. do they possess the equilibrium
distribution?

\medskip
In the frame of
reference moving with the flow, the equilibrium distribution is Maxwellian.  We
demonstrate this by showing that the electrons and photons constitute
a canonical ensemble in equilibrium.  In this system the
electrons compose a subsystem in thermal contact with a heat bath made up of
the photons.  To constitute a canonical ensemble, the particles must
satisfy two requirements.  First, the number of particles in the
subsystem must be much smaller than the number of particles in the
heat bath.  Second, the sum of the energies of the system and the heat
bath must be constant (see, e.g., \citeNP{ll80}).
Since $n_e \ll n_x E_d^1$ and energy is conserved in
electron-photon scattering, both of these conditions are satisfied.
Thus, the probability that an electron has an energy $(p^2+m_e^2
c^4)^{1/2}$ 
is a Maxwell-Boltzmann distribution.  

\medskip
The electrons relax to
equilibrium by electron-photon scattering in a time $\sim t_{e\gamma}$.
Note that there is no requirement that the {\it photon} distribution
be thermal or that there be significant electron-proton
scattering for the electrons to reach equilibrium.
Since $t_{e\gamma} \ll t_{esc}$, the electrons in the outflow have an
equilibrium distribution, except in the tail of the Maxwellian;
the electrons do not remain in the line-forming region long enough to
scatter a sufficient number of times to populate the tail.  Since the
number of electrons in the tail is small, however, they do not have a
significant impact on the properties of the emerging radiation.
Therefore, in our calculations we assume an equilibrium distribution at
all electron momenta.

\medskip
The electron temperature at equilibrium (i.e. the Compton temperature) $T_C$
is equal to the temperature of the photons $T_\gamma \equiv
(dS/dU)^{-1}$, where $S$ and $U$ are the entropy and energy of the
photons, respectively.  The Compton
temperature can be determined microphysically by summing the
energy transferred from the radiation 
to the plasma in each scatter; at the Compton temperature
this sum is zero.  Using the single scattering
approximation, Lamb, Wang, and Wasserman (1990\nocite{lww90})
calculate $T_C$ in the limits $b,\beta \ll 1$ for a static plasma and
first harmonic scattering only.

\medskip
We adapt their model to a dynamic plasma by finding the Compton
temperature $T_C^{\rm f}$, where the superscript f indicates
quantities measured in the frame of reference moving with the flow.
We then boost the momenta distribution back to the lab frame.  
From eq. (\ref{angledist}), eq. (\ref{ngammaE}), and the Lorentz
invariance of $E^{-2} 
n_x$ (see, e.g. \citeNP{ll75}),

\begin{equation}
n_x^{\rm f}(E^{\rm f},\mu^{\rm f})={\eta(\mu^{\rm f}-\mu_o^{\rm
f}) \over \bigg 
[\gamma_F (1+\beta_F \mu^{\rm f}) \bigg ]^{s+2}\ E^s}\ ,
\label{ngf}
\end{equation}

\noindent 
Substituting (\ref{ngf}) into the formula of Lamb, Wang,
and Wasserman (1990),

\begin{equation}
k_B T_C^{\rm f} = {E_B \over 10} {\int_{-1}^{+1} {(2+7\mu^{\rm f 2}+5\mu^{\rm f
4})\ n^f_x(E_B,\mu^{\rm f})\  
d\mu^{\rm f}} \over 
\int_{-1}^{+1} {[1+(2+s)\mu^{\rm f 2}+(s-3)\mu^{\rm f 4}]\ n_x^{\rm
f}(E_B,\mu^{\rm f})\ d\mu^{\rm f}}}\ ,
\label{tclww}
\end{equation}

\noindent
and using eq. (\ref{bfapprox2}) for the velocity profile gives $k_B
T_C^{\rm f} \approx 0.25 E_B$ for all $z$.  We use this value of the
Compton temperature throughout the present work.  
For a Maxwellian distribution with temperature
$k_B T^{\rm f} \ll \gamma_F^{-2}\ m_e c^2$ in the frame moving with
the flow, the {\it lab} frame distribution is approximately
Maxwellian, with the peak at $p=p_F \equiv \gamma_F\ \beta_F\ m_e c$
and temperature $T_e=\gamma_F^2 T_e^{\rm f}$.  The approximation is a
good one near the peak, but breaks down in the tail.  However, as we
point out above, the shape of the distribution in the tail does not  
have a significant impact on the properties of the emerging radiation.

\medskip
Monte Carlo calculations by Lamb, Wang and Wasserman
(1990\nocite{lww90}) and Isenberg, Lamb, and Wang (1997\nocite{ilw97})
show that when higher harmonics, multiple scattering, and the geometry
of the line-forming region are included, $T_C$ varies from the
analytic value (\ref{tclww}) but is always between $0.2 E_B$ and $0.7
E_B$.  For the line energies observed by {\it Ginga}, the full-width
half-maximum of a line formed in an outflow is relatively insensitive
to the temperature since the contribution of thermal broadening to the
the line width is small.  This contribution is just the Doppler width
$E_d^1=3.2$ keV for $B_{12}=1.7$, $T_k^{\rm f}=5.0$, and
$\beta_F=0.5$.  By comparison, the broadening of the line due to the
variation in the field with altitude is $\Delta E_B \equiv |dE_B/dz|\
r_h = 6.0$ keV for the same field strength.  Therefore the total line
width is $\approx [ (\Delta E_B)^2 + (E_d^1)^2]^{1/2} = 6.8\ {\rm
keV}\ \approx \Delta E_B$.  We show in Section \ref{sprofiles} that the
optical depth, and therefore the {\it equivalent} width, is also
insensitive to the temperature.  Therefore, eq. (\ref{tclww}) is
sufficiently accurate for determining the properties of the emerging
spectrum.

%%%%%%%%%%%%%%%%%%%%%%%%%%%%%%%%%%%%%%%%%%%%%%%%%%%%%%%%%%%%%%%%%%%
%                                                                 %
%     Plasma Density Profile                                      %
%                                                                 %
%%%%%%%%%%%%%%%%%%%%%%%%%%%%%%%%%%%%%%%%%%%%%%%%%%%%%%%%%%%%%%%%%%%

\subsection{Plasma Density Profile}
\label{spdens}

Conservation of matter in an outflow requires that in the absence of
plasma sources or sinks, $n_e(z)\ 
\beta_F(z)\ \tilde{r}^2(z)$ is a constant.  The conservation of magnetic
flux determines the radius of the outflow $r(z)$, according to
eq. (\ref{flare}).  Therefore, if the plasma at the surface has
density $n_{eo}$, the density at altitude $z$ is (\citeNP{miller91})

\begin{equation}
n_e(z)=n_{eo}\ {\beta_{Fo} \over \beta_F(z)}\ {B(z) \over B_o}\ .
\label{nez}
\end{equation}

%%%%%%%%%%%%%%%%%%%%%%%%%%%%%%%%%%%%%%%%%%%%%%%%%%%%%%%%%%%%%%%%%%%
%                                                                 %
%     Scattering Kinematics                                       %
%                                                                 %
%%%%%%%%%%%%%%%%%%%%%%%%%%%%%%%%%%%%%%%%%%%%%%%%%%%%%%%%%%%%%%%%%%%

\subsection{Scattering Kinematics}
\label{skine}

A resonant scatter may be thought of as an absorption of a photon by
an electron, which excites the electron to a higher Landau level,
followed by the emission of one or more photons.  The absorption
conserves energy and momentum along the magnetic field; the portion of
the photon's momentum perpendicular to the field that is not
transferred to the electron is absorbed by the field.  Combining the
conservation laws for a transition from the 0 to $n$th Landau state,
and ignoring the finite natural line width, yield the resonant
condition ($c=m_e=1$ throughout this section)

\begin{equation}
1+E^r - \sqrt{1+2nb+\mu^{r2} E^{r2}}=0\ ,
\label{econs}
\end{equation}

\noindent where the superscript $r$ indicates quantities measured in
the frame of reference
where the electron is at rest prior to scattering.  Solving
eq. (\ref{econs}) for the rest frame resonant photon energy yields

\begin{equation}
E_n^r={2nb \over 1 + \sqrt{1+2nb\ (1-\mu^{r2})}}\ .
\end{equation}

\noindent Transforming to the lab frame (the frame of reference where the
magnetic dipole is at rest),

\begin{equation}
E_n = {E_n^r \over \gamma\ (1-\beta \mu)}\ .
\label{eblab}
\end{equation}

\medskip
We plot the first harmonic lab frame resonant energy as a function
of electron momentum $p$ in figure \ref{fecons}.  As the figure shows,
for a given photon energy and $|\mu|<1$ there are two roots for the
electron momentum; $p_-$
and $p_+$ denote the lower and upper roots respectively.  Physical
solutions to the resonant condition do not exist above the cutoff
energy 

\begin{equation}
E_c={\sqrt{1+2nb}-1 \over \sqrt{1-\mu^2}}
\end{equation}

\noindent  (\citeNP{dv78}; \citeNP{hd91}; \citeNP{wwl93}).  To
understand the physical origin of the two momentum roots and the
cutoff energy, consider a frame of reference
(indicated with a prime) which is moving with a velocity
$\beta = \mu$ with respect to the lab frame.  In the primed frame, the
photon is moving 
perpendicular to the magnetic field ($\mu^\prime=0$), since

\begin{equation}
\mu^\prime={\mu - \beta \over 1 - \mu \beta}
\end{equation}

\noindent In this frame the resonant condition is symmetric with
respect to reflection in the plane perpendicular to the field.  Thus
there must be two equal and opposite solutions for $p^\prime$;
i.e. the $E_n^\prime$ vs. $p^\prime$ curve is symmetric about
$p^\prime=0$ and $p_+^\prime=-p_-^\prime$, as shown in figure
\ref{fecons}.  For non-zero electron momentum the photon appears
redshifted in the electron rest frame.  But at $p^\prime=0$,
$\mu^r=\mu^\prime=0$, there is no redshift, and the resonant energy
has its maximum value $E_n^\prime=E_c^\prime=E_n^r=\sqrt{1+2nb}-1$.
Boosting $p_\pm^\prime$ and $E_c^\prime$ back to the lab frame gives
the momentum roots and cutoff energy when $\mu \neq 0$.

\medskip
Figure \ref{fecons} illustrates that the scattering kinematics in an
outflow are noticeably different from those in a static plasma.  For a
given $\mu$ the line center energy $E_{ctr}$ is equal to the resonant
energy corresponding to $\beta=\beta_F$.  In a static plasma
($\beta_F=0$), $E_{ctr} \approx E_B$ for all $\mu$.  The cutoff energy
gives the scattering cross section a strong asymmetry near $\mu=0$ (\citeNP{lamb89}) but
plays little role at larger $\mu$ where the cutoff energy is far above
the line center.  At these angles, the $p_-$ root dominates the
scattering.  However in an outflow with, for example, $\beta_F=1/2$,
there is a large variation in the line center energy with photon
direction, as expected from eq. (\ref{eblab}).  The effects of the
cutoff energy are the most dramatic at $\mu=\beta_F$, where
$E_{ctr} = E_c$.  $p_-$ dominates the scattering at
larger values of 
$\mu$ while $p_+$ is dominant at smaller ones.

%%%%%%%%%%%%%%%%%%%%%%%%%%%%%%%%%%%%%%%%%%%%%%%%%%%%%%%%%%%%%%%%%%%
%                                                                 %
%      Scattering Profiles                                        %
%                                                                 %
%%%%%%%%%%%%%%%%%%%%%%%%%%%%%%%%%%%%%%%%%%%%%%%%%%%%%%%%%%%%%%%%%%%

\subsection{Scattering Profile}
\label{sprofiles}

In calculating the scattering profiles we assume that the vacuum
contribution to the dielectric tensor is much larger than the plasma
contribution.  This is true provided

\begin{equation}
{w \over \delta} =4 \times 10^{-5}\ n_{e,17}\ B_{12}^{-4} \ll \left | 1 - \left( E_B \over E \right )^2 \right |,
\label{wodel}
\end{equation}

\noindent
where $w \equiv (\hbar \omega_p/E_B)^2$ is the plasma frequency
parameter, $\omega_p$ is the plasma frequency, and $\delta$ is the
magnetic vacuum polarization parameter (see \citeNP{adler71};
\citeNP{wws88}).  For the
fields and densities considered in the present work, eq. (\ref{wodel})
breaks down above $z \sim R_*$ where the field strength
becomes small.  However, the condition is satisfied near the stellar
surface where most of the scattering takes place.  
     This condition also breaks down very near the line center, 
     where $E \approx E_B$.  However, the effect on the emerging
     spectrum is negligible since the line properties are determined 
     primarily by the scattering profile in the line wings (see below). 
  
The scattering cross sections we use are valid in the limit

\begin{equation}
\left ( {E^r \over m_e c^2} \right )^2 {1\over b} \ll 1\ ,
\label{tsmall}
\end{equation}

\noindent
which holds throughout the present work.  
In this limit the
scattering cross section approaches the classical magnetic Compton
cross section.  When evaluated in the pre-scattered
electron's rest frame and averaged over the azimuthal angle
(appropriate for azimuthally symmetric line-forming regions such as
those studied here) 
this cross section is given by (\citeNP{clr71}; \citeNP{herold79};
\citeNP{ventura79}; \citeNP{ws80}; \citeNP{wws88};
\citeNP{lww90}; \citeNP{hd91}; \citeNP{graziani93}; 
\citeNP{ilw97}): 

\begin{equation}
{d\sigma^r \over d\Omega^r_s} = {d\sigma^r_0 \over d\Omega^r_s}
+ {d\sigma^r_1 \over d\Omega^r_s}\ .
\label{ClassCross}
\end{equation}

\noindent
The first term on the right is the continuum part of the cross section

\begin{equation}
{d\sigma^r_0 \over d\Omega^r_s} = {3 \over 16
\pi} \sigma_T \times \left \{ \
\begin{array}{lc}
2 \sin^2 \theta^r \sin^2 \theta^r_s + \left( {E^r \over
   E^r + E_B} \right)^2 \mu^{r2} \mu_s^{r2} & (\| \rightarrow \|) \\
\left( {E^r \over E^r + E_B} \right)^2 \mu^{r2} & (\| \rightarrow
   \perp) \\
\left( {E^r \over E^r + E_B} \right)^2 \mu_s^{r2} & (\perp \rightarrow
   \|) \\
\left( {E^r \over E^r + E_B} \right)^2 & (\perp \rightarrow
   \perp)
\end{array}
\right. \ .
\label{NonResCross}
\end{equation}

\noindent
The symbols at the right of eq. (\ref{NonResCross}) indicate the polarization
mode, parallel or perpendicular, of the initial and scattered photon.
The resonant part of the cross section is 

\begin{eqnarray}
{d\sigma^r_1 \over d\Omega^r_s} & = &
{9 \over 64} {\sigma_T m_e c^2 \over \alpha} 
\left [   {4 E^{r3} \over E_B (E^r+E_B)^2} \right ]
\times \nonumber \\
& & {\Gamma_1/2\pi \over (E^r-E^r_1)^2 + (\Gamma_1/2)^2} \times
\left \{ 
\begin{array}
{cc}
\mu^{r2} \mu_s^{r2} & (\| \rightarrow \|)  \\
\mu^{r2} & (\| \rightarrow \perp)  \\
\mu_s^{r2} & (\perp \rightarrow \|)  \\
1 & (\perp \rightarrow \perp)  \\
\end{array}
\right. \ . 
\label{ResCross}
\end{eqnarray}

\noindent
The corresponding polarization-averaged cross sections are calculated
by summing eqs. (\ref{NonResCross}) and (\ref{ResCross}) over the
final polarization states and averaging over the 
initial. 

\medskip
The contribution of $n>1$ harmonic scattering to the scattering profile can
be calculated approximately by treating higher harmonic scattering as the
absorption of a photon followed by the emission of one or more
photons.  Then the total rest frame cross section

\begin{equation}
\sigma^r \approx \sum_{n=0}^\infty \sigma^r_n\ ,
\label{SigTot}
\end{equation}

\noindent where

\begin{equation}
\sigma_0^r=\sum_{\|,\perp}{\int{d\Omega_s^r {d\sigma^r_0 \over
d\Omega^r_s}}}\ ,\ \ \ \ \
\sigma_1^r=\sum_{\|,\perp}\int{d\Omega_s^r {d\sigma^r_1 \over
d\Omega^r_s}}\ 
\end{equation} 

\noindent (the sums are over the {\it final} polarization states), and

\begin{eqnarray}
\sigma_{n>1}^r & = & {3 \over 4}
{\pi m_e c^2 \sigma_T \over \alpha}\ b^{n-1}\ {(n^2/2)^{n-1} \over
(n-1)!} \times \nonumber \\
& & {\Gamma_n/2\pi \over (E^r-E^r_n)^2 + (\Gamma_n/2)^2}\ (1 - \mu^{r2})^{n-1}
 \times
\left \{ 
\begin{array}
{cc}
\mu^{r2} & (\|)  \\
1 & (\perp)
\end{array}
\right. 
\label{sigrest}
\end{eqnarray} 

\noindent
is the total cross section for absorption at the $n$th
harmonic (Daugherty and Ventura 1977; Fenimore et al. 1988).
The approximation (\ref{SigTot}) is valid provided
the line forming region is not optically thick in the
line wings of the first harmonic (see below; also \citeNP{ws80};
\citeNP{lamb89}; \citeNP{ilw97}), i.e.

\begin{equation}
\tau_1 \lesssim 1/a\ ,
\label{atausmall}
\end{equation}

\noindent
where $\tau_1$  is the polarization and frequency-averaged first
harmonic optical depth for photons moving parallel to the magnetic
field (see below), and $a$ is given by eq. (\ref{natline}).
Eq. (\ref{atausmall})
holds for all cases we study in the present work. 

\medskip
From the Lorentz invariance of the optical depth (see, e.g.,
\citeNP{rl79}), the lab frame cross section $\sigma_n$ is
related to $\sigma_n^r$ by

\begin{equation}
\sigma_n = (1-\beta\mu)\,\sigma_n^r\ .
\label{siglab}
\end{equation} 

\noindent
We average the lab frame cross sections over the electron momenta
distribution and 
divide by the Thomson cross section to obtain the scattering profile
for the $n$th harmonic

\begin{equation}
\phi_n(E,\Omega) \equiv {1 \over \sigma_T}\ \int dp f(p) \sigma_n\ .
\label{NScatProfile}
\end{equation}

\noindent The line can be divided into the line core ($|x_n/\mu| \ll
1$) and the line wings ($|x_n/\mu| \gg 1$), where $x_n \equiv (E -
E_n)/E_d^n$ is the dimensionless frequency shift (Wasserman and
Salpeter 1980).   In the line core, the electron distribution
dominates the profile so that $\phi_n \propto f(x_n^2/\vert \mu
\vert)$.  In the wings, the tail of the Lorentzian distribution
dominates so that $\phi_n \propto  a x_n^{-2}$.  We refer to the wing
at energies below the line center as the red wing and the wing at
energies above the line center as the blue wing.  Wasserman and
Salpeter (1980) show
that for the first harmonic, the core-wing boundary appears at
$|x_1/\mu| \approx 2.62 - 0.19 \ln(100\ a/\mu)\ $.

\medskip
The optical depth at the $n$th harmonic is equal to 

\begin{equation}
\tau_n(E,\mu)=\sigma_T \int{\phi_n(E,\mu)}\ {n_e(z) dz \over |\mu|}\ .
\end{equation}

\noindent The first harmonic optical depth along
the line of sight with the shortest escape path characterizes the
radiation transfer.  In the outflow
model this line of sight is along the field, due to the change in the
cyclotron energy with altitude: a photon
can escape the line core by diffusing to an altitude where it is in
the wings.  If the
characteristic distance a photon climbs before escaping the line core

\begin{equation}
\Delta z_{esc} \equiv \left |{E_d^1 \over dE_B/dz} \right |_{z=0} \ll
R_*, 
\end{equation}

\noindent then the energy averaged first harmonic optical depth
along the field is 

\begin{equation}
\tau_1 \approx \sigma_T \left | \int{ 
\phi_n(E,1)\ {dE \over E_d^1}}\right |_{z=0} n_{eo}\ \Delta z_{esc} \ .
\end{equation}

\noindent
Averaging over the photon polarization and
adopting the velocity profile from
the model of MT and
a Gaussian electron momentum distribution yields

\begin{equation}
\tau_1 \approx 82\ {n_{eo,17}\ r_{h,5} \over B_{o,12} \left(1-4.5\ 
r_h / R_* \right )}\ .
\label{tau1}
\end{equation}

\medskip
For $r_h / R_*= 0.1$, eq. (\ref{tau1}) gives $\tau_1 \approx 149\
n_{eo,17}\ r_{h,5}\ B_{o,12}^{-1}$,
 and the condition eq. (\ref{atausmall}) implies
 $n_{eo,17}\lesssim 4\,\,r_{h,5}^{-1}\,\,T_{k}^{1/2}$.
Comparing eq. (\ref{tau1}) for $\tau_1$ to the
corresponding optical depth for a static slab line-forming region is
instructive.  In a slab with a uniform magnetic field parallel to the
slab normal, the optical depth along the field is $\tau_1 = 1500\
N_{e,22}\ B_{12}^{-1}\ T_k^{-1/2}$ (\citeNP{lww90}), where $N_{e,22}$ is
the column depth in units of $10^{22}\ {\rm cm^{-2}}$.  The optical
depth of an outflow is about an order of magnitude smaller than that
of a slab with a comparable column depth, i.e., a slab whose column
depth is equal to the {\it radial} column depth $n_{eo}r_h$ of the
outflow.  The optical depth is smaller for the outflow because the
distance a photon needs to travel to escape the line core is smaller
than $r_h$; it can escape by climbing to an altitude where the
cyclotron energy is less than the photon energy by more than a Doppler
width.  Note that in the outflow, the optical depth is independent of
the electron temperature, in contrast to the slab where $\tau_1 \sim
T_k^{-1/2}$.  The reason for this is that in the outflow the decrease in
$d\tau_1/dz$ with increasing temperature is offset by the greater
distance a photon needs to climb to escape the line core.

%%%%%%%%%%%%%%%%%%%%%%%%%%%%%%%%%%%%%%%%%%%%%%%%%%%%%%%%%%%%%%%%%%%
%                                                                 %
%      Gravitational Red Shift                                    %
%                                                                 %
%%%%%%%%%%%%%%%%%%%%%%%%%%%%%%%%%%%%%%%%%%%%%%%%%%%%%%%%%%%%%%%%%%%

\subsection{Gravitational Red Shift}
\label{srshift}

A photon that travels from the stellar surface ($z=0$) to an observer
at $z=\infty$ experiences a gravitational red shift

\begin{equation}
{E(0)-E(\infty) \over E(0)}={GM_* \over c^2 R_*} = 0.2
\left ({M_* \over M_{ch}} \right ) \left ({R_* \over 10^6\ {\rm cm}}
\right )^{-1},
\end{equation}

\noindent
where $M_*$ is the stellar mass and $M_{ch}=2.8 \times 10^{33}$ g is
the Chandrasekhar mass.  Because the red shift changes the photon
energy by only $\approx$ 20\%, we do not include it in our calculations.
The red shift reduces the cyclotron line width as well as the
energy of individual photons, so the red shift narrows
the lines.  Since we wish to determine whether narrow lines can be
formed in an outflow, ignoring the red shift is a {\it conservative}
assumption.  

%%%%%%%%%%%%%%%%%%%%%%%%%%%%%%%%%%%%%%%%%%%%%%%%%%%%%%%%%%%%%%%%%%%
%                                                                 %
%      Electron Velocity Profiles                                 %
%                                                                 %
%%%%%%%%%%%%%%%%%%%%%%%%%%%%%%%%%%%%%%%%%%%%%%%%%%%%%%%%%%%%%%%%%%%

\section{Electron Velocity Profiles}
\label{bprof}

The net force on a particle in an outflow
is the sum of the radiation and gravitational forces, ${\bf
F}={\bf F}_{rad}+{\bf F}_g$.  For the luminosities appropriate for a
gamma-ray burst in the galactic corona, the radiation force is much
larger than the gravitational force, so that ${\bf F} \approx {\bf
F}_{rad}$.  Since the electrons are restricted to motion along the
field lines, only the component of the radiation force parallel to the
magnetic field is of interest.  In Section \ref{sMT} we use the
approximate, analytic model of MT to calculate the radiation force
along the field line.  We then integrate the equations of motion to
determine the velocity profile.  In Section \ref{selfcon} we use a
Monte Carlo radiative transfer code to determine the velocity profile
self-consistently.

%%%%%%%%%%%%%%%%%%%%%%%%%%%%%%%%%%%%%%%%%%%%%%%%%%%%%%%%%%%%%%%%%%%
%                                                                 %
%      Analytic Model of Mitrofanov and Tsygan                    %
%                                                                 %
%%%%%%%%%%%%%%%%%%%%%%%%%%%%%%%%%%%%%%%%%%%%%%%%%%%%%%%%%%%%%%%%%%%

\subsection{Analytic Model of Mitrofanov and Tsygan}
\label{sMT}

MT calculate the radiation force in an outflow using a single
scattering approximation.  They show that when first harmonic
scattering dominates the radiation transfer, and $\Gamma_r \rightarrow
0$, the component of the radiation force along the field line is 

\begin{eqnarray}
F_{rad} & = &\pi^2 \sigma_T\ {E_B^2 \over \Gamma_r}\ E_B \int{(\mu - \beta)\ \left
[1+\left({\mu-\beta \over 1 - \beta \mu} \right )^2 + P_l {1-\mu^2
\over \gamma^2 (1-\beta \mu)^2} \right ] \times} \nonumber \\
& & {1 \over 1 - \beta \mu} \
n_x \left ({E_B \over \gamma (1-\beta \mu)},\ \mu \right) d\mu \ ,
\label{fradMT}
\end{eqnarray}

\noindent where $P_l$ is the degree of linear polarization, i.e. the
fraction of perpendicular mode photons minus the fraction of parallel
mode photons.  Because eq. (\ref{fradMT}) is linear in the scattering
cross section, the radiation force for the polarization averaged cross
section is the same as the radiation force for polarized cross
sections with $P_l=0$.  Substituting the power law photon spectrum we
use in the present work [eqs. (\ref{ngammaEO}) and (\ref{ngammaE})]
into the radiation force gives

\begin{equation}
F_{rad} = {\pi \sigma_T \over c}\ {E_B^2 \over \Gamma_r}\ {L_x
\gamma^s E_B^{1-s} \over r_h^2\ \xi(s)}\ I_1\ ,
\label{frad2}
\end{equation}

\noindent where

\begin{equation}
I_1 \equiv \int_{\mu_o}^1{(\mu-\beta)\ \left
[1+\left({\mu-\beta \over 1 - \beta \mu} \right )^2 + P_l {1-\mu^2
\over \gamma^2 (1-\beta \mu)^2} \right ]\ (1-\beta \mu)^{s-1}\ Q(\mu)
d\mu} \ .
\label{int1}
\end{equation}

\noindent  Eq. (\ref{frad2}) is equivalent to

\begin{eqnarray}
{d\gamma \over d\tilde{z}} & = & r_h \left \{ 
\begin{array}{c} (m_p c^2)^{-1} \\ (m_e
c^2)^{-1} \end{array} \right \} F_{rad} \nonumber \\ 
& = & \left \{ \begin{array}{c} 2.4 \times
10^5 \\ 4.5 \times 10^{8} \end{array}
 \right \} \ L_{x,40}\ \gamma^s \left ({E_B \over 1\ {\rm keV}} \right )^{1-s}
r_{h,5}^{-1} \left ({\xi(s) \over 1000\ {\rm keV}^{2-s}} \right )^{-1}
I_1\ ,
\label{dgdz}
\end{eqnarray}

\noindent where the upper expressions in curly brackets are for an
electron-proton plasma and the lower expressions are for an
electron-positron plasma.

\medskip
By integrating eq. (\ref{dgdz}), we calculate the velocity as a
function of altitude for an electron in an electron-proton plasma,
injected at the origin with $\beta=0$.  We use a $1/E$ photon
spectrum with $P_l=0$.  We plot the result in Figure \ref{b_vs_z}.  The
curve shows three distinct regimes which MT call free acceleration, kinematic
restriction, and energetic restriction.  At any altitude $z$ there
exists a velocity $\beta_{(F=0)}$ where the radiation force is equal to
zero.  Near the surface, $\beta_{(F=0)} \approx 0.5$, which corresponds
to $\gamma_{(F=0)}\approx 1.15$.  In the free acceleration regime
$\beta<\beta_{(F=0)}$ and the electron accelerates rapidly.  

\medskip
When
$\beta$ reaches $\beta_{(F=0)}$ the electron enters the regime of kinematic
restriction.  The distance it travels before reaching this regime is

\begin{equation}
\tilde{z}_{KR} \sim {\Delta \gamma \over d\gamma/d\tilde{z}} \sim  \left \{
\begin{array}{c} 10^{-6}\\ 10^{-9} \end{array} \right \}\ 
L_{x,40}^{-1} \left ({E_B \over 1\ {\rm keV}} \right )^{s-1} r_{h,5}
\left ({\xi(s) \over 1000\ {\rm keV}^{2-s} } \right )\ .
\label{zkr}
\end{equation}

\noindent In the kinematic restriction $\beta \approx \beta_{(F=0)}$.
The electron requires a small radiation force to overcome the force of
gravity and keep pace with the increase of $\beta_{(F=0)}$ with altitude.
However the difference between the actual velocity and
the zero force velocity is $\sim$ one part in $10^6$.  

\medskip
As the electron continues to climb, the radiation becomes more diffuse
and eventually carries too little energy to maintain the particle
at the zero force velocity; i.e.

\begin{equation}
{d\gamma_{(F=0)} \over d\tilde{z}}\ >\ {d\gamma \over d\tilde{z}}\ .
\label{erstart}
\end{equation}

\noindent
The particle enters the energetic restriction regime and
the Lorentz factor levels off to $\gamma=\gamma_{max}$.  For $z \gg 1$ 
and $s=1$, eq. (\ref{int1}) becomes

\begin{equation}
I_1\ \approx\ {1 \over 2 \tilde{z}^2} \left [ {1 \over
\gamma^2} - {3 \over 2 \tilde{z}^2} \right ]\ \leq\ {1 \over
2 \tilde{z}^2 \gamma^2}\ ,
\label{i1bigz}
\end{equation}

\noindent so $\gamma_{(F=0)} \approx (2/3)^{1/2} \tilde{z}$.
Substituting $\gamma \sim \tilde{z}$ and eq. (\ref{i1bigz}) into
eq. (\ref{erstart}), the energetic restriction begins at  

\begin{equation}
\tilde{z}_{ER} \sim \gamma_{max} \sim \left \{ \begin{array}{c} 100\\ 1000
\end{array} \right \}\ L_{x,40}^{1/3}\ r_{h,5}^{-1/3}\ .
\label{zer}
\end{equation}

\noindent
We caution that the discussion of the energetic restriction is
highly speculative.  Since the cyclotron energy in this regime is
below burst detector thresholds, we do not know the relevant part of
the photon spectrum observationally.

\medskip
Figure \ref{b_vs_z} and eqs. (\ref{zkr}) and (\ref{zer}) clearly
show that the electron is in the kinematic restriction throughout the
region of interest for line formation, except for about a millimeter
of free acceleration near the surface.  Consequently, we assume
$F_{rad}=0$ for the remainder of the present work.

\medskip
Eq. (\ref{dgdz}) gives the acceleration of an {\it individual}
electron.  For a group of electrons whose velocities are distributed
thermally about the flow velocity, the
radiation force should be averaged over the velocity distribution.   
Setting the average force to zero and solving for the flow velocity
yields a velocity almost indistinguishable from the zero force
velocity of an individual electron, i.e. $\beta_F \approx
\beta_{(F=0)}$.  There are two reasons for this. 
First, the corrections to the velocity due to temperature are $\sim
k_B T_e / m_e c^2 \ll \beta_{(F=0)}$.  Second, the
Maxwell-Boltzmann distribution is symmetric, so the corrections due to
electrons with velocities on each side of the bulk flow velocity tend
to cancel. 

\medskip
Figure \ref{bf_vs_z.s} shows the zero force velocity profile given by
eq. (\ref{frad2}) with $P_l=0$ and $s=$ 0.5, 1.0, and 2.0.  For
comparison, we also plot the approximate velocity profile of
eq. (\ref{bfapprox2}).  For $s <\ \approx 1.5$, the flow velocity is
larger than the approximate value.  This is due to the $1+\mu^2$
dependence of the scattering cross section, which the approximate
calculation does not take into account.  Photons moving at large
$\mu$, which carry more momentum along the field line, are more likely
to scatter, increasing the radiation force and hence the flow
velocity.  As the spectral index increases, there are fewer photons
with large energies.  Consequently $F_{rad}$ is smaller and $\beta_F$
decreases.

\medskip
Figure \ref{bf_vs_z.pol} shows the velocity profile for $s=1$ and
$P_l=\ -1$, 0, and 1.  In perpendicular mode,
the scattering cross section is independent of $\mu$ and the velocity
profile 
is the same as the approximate one.  As the fraction of
parallel mode photons increases, photons at small angles to the field
are more likely to scatter and the flow velocity increases.

%%%%%%%%%%%%%%%%%%%%%%%%%%%%%%%%%%%%%%%%%%%%%%%%%%%%%%%%%%%%%%%%%%%
%                                                                 %
%      Self-Consistent Monte Carlo Model                          %
%                                                                 %
%%%%%%%%%%%%%%%%%%%%%%%%%%%%%%%%%%%%%%%%%%%%%%%%%%%%%%%%%%%%%%%%%%%

\subsection{Self-Consistent Monte Carlo Calculation}
\label{selfcon}

We calculate self-consistent velocity profiles by iterating
calculations of the radiation transfer and the hydrodynamics.  The
procedure is conceptually similar 
to the one Wang and Frank (1981\nocite{wf81}) use to determine the
velocity profile and radiation energy density in the accretion column
of an accretion-powered pulsar.  For the
radiation transfer, we have adapted a Monte Carlo code written for
static, slab line-forming regions (\citeNP{wws88}; \citeNP{wang89};
\citeNP{ilw97}) to the geometry and physical conditions in an outflow.
The code calculates the radiation force, which we integrate to
determine the velocity profile for the next iteration.  We repeat the
process until the radiation force is zero everywhere, within the
statistical uncertainty of the Monte Carlo code.  

\medskip
In addition to the physical processes included in the analytic model of
MT, the Monte Carlo calculation permits
multiple scattering, higher harmonics, and a finite natural line width.  
The analytic model assumes a disk photon source with electrons on the
cylinder axis, or equivalently, a point photon source with electrons
anywhere in the outflow.  The Monte Carlo code permits electrons
anywhere in the outflow and either a point or a disk photon source.  

\medskip
Figure \ref{bf_vs_z.goh} illustrates the effect of these enhancements
on the velocity profile.  The figure shows the velocity profile for a
1/E photon spectrum injected into a line-forming region with
$r_h=0.1\ R_*$, $B_{o,12}=1.7$, and $n_{eo}=10^{17}\ {\rm cm^{-3}}$.  We use
polarization averaged cross sections in all simulations, except where
otherwise indicated.  As the figure shows, for a Monte Carlo
calculation with first harmonic scattering only, a point photon
source, and $\Gamma_r=0$, the self-consistent velocity profile is
very similar to the profile calculated from MT's analytic model.  The
two calculations agree within about 10\% at the stellar surface.  To understand
the difference, consider the optical depth of a region which is small
enough that the magnetic field and flow velocity do not change
significantly.  The fraction of photons with energy $E$ and
orientation $\mu$ which scatter in this region is $f_s \approx
1-\exp[-\tau(E,\mu)]$.  Assuming that first harmonic scattering
dominates the radiation transfer, expanding in $\tau$, and averaging
over energy yields, in the frame of reference moving with the flow,

\begin{eqnarray}
f_s^{\rm f} & \approx & 1-e^{-3/4\ \tau_1 (1+\mu^{\rm f2})}
\nonumber \\ & = & {3 \over 4}
\tau_1 (1+\mu^{\rm f2}) \left [1-{3\over 8} \tau_1 (1+\mu^{\rm f2}) +
O(\tau_1^2) \right ]
\label{fscat}
\end{eqnarray}

\noindent where $\tau_1$ is the energy {\it and angle} averaged
optical depth.  When the optical depth is very small ($\tau_1 \ll 1$),
$f_s$ is proportional to the cross section, consistent with the single
scattering approximation used by MT.  Since $F_{rad} \propto
\tau_1$, and
$F_{rad}=0$ at the flow velocity, the velocity profile is not
sensitive to $\tau_1$ at very small optical depths.  However, as the
optical depth increases, the higher order terms in eq. (\ref{fscat})
become significant.  Consequently, photons traveling at large angles to
the field are more likely to scatter, reducing $\beta_F$.

\medskip
The finite natural line width reduces the flow velocity even further.
We attribute this to scattering of photons in the line wings.  Wing
scattering has the greatest impact at $\mu=0$ where there is no
thermal broadening of the cyclotron line.  Consequently, the finite
natural line width increases the number of scatters of photons at
large angles to the magnetic field, decreasing $\beta_F$.  

\medskip
When photons are injected semi-isotropically at the origin of the hot
spot, the mean distance from the injection point to the walls of the
cylinder is $\pi/2\ r_h \approx 1.57\ r_h$.  When they are injected
uniformly across the hot spot, this distance falls slightly, to $4/3\
r_h \approx 1.33\ r_h$, resulting in a proportional drop in mean
optical depth along the line of sight.  Since the dependence of the
velocity profile on optical depth is second order and the change in
optical depth is small, the method of injection does not have a
significant impact on the profile, as Figure \ref{bf_vs_z.goh} shows.

\medskip
When higher harmonic scattering is added to the calculation, the flow
velocity drops further.  This is clear from the $(1-\mu^2)^{n-1}$
dependence of the scattering cross sections.  Photons with small $\mu$
are more likely to scatter at higher harmonics, lowering the radiation
force.

\medskip
We illustrate the effect of the optical depth on the velocity profile
in greater detail in Figure \ref{bf_vs_z.edens}.  The figure shows velocity
profiles for a line-forming region with $r_h=0.05\ R_*$ and
$B_{o,12}=1.7$.  Continuum and first three harmonic scattering with
finite natural line width are included in the calculation.  The
injected photon spectrum is the best fit of a double power law to the
continuum spectrum of GB880205, $n_x(E) \propto E^{-s}$,
where $s = \alpha_1 = 0.8479$ for $E<E_{break}=107.8$ keV and $s =
\alpha_2 = 1.199$ for
$E>E_{break}$ (\citeNP{pefpc}; see also \citeNP{fenimore88};
\citeNP{wang89}).  As $n_{eo}$ increases from $10^{16}\ {\rm cm^{-3}}$
to $10^{17}\ {\rm cm^{-3}}$, the flow velocity decreases, as expected
from eq. (\ref{fscat}).  However, as the plasma density increases
further, to $10^{18}\ {\rm cm^{-3}}$, the trend reverses.  This is
because at the larger plasma density there is significant angular
redistribution of the photons; eq. (\ref{fscat}) does not take this
redistribution into account.  Because of the motion of the plasma,
many photons are scattered to large values of $\mu$ (see Figures
\ref{spectra.basic} and \ref{spectra.edens}, below).  Further
scattering of these photons increases $F_{rad}$ and $\beta_F$.

%%%%%%%%%%%%%%%%%%%%%%%%%%%%%%%%%%%%%%%%%%%%%%%%%%%%%%%%%%%%%%%%%%%
%                                                                 %
%      Monte Carlo Spectra                                        %
%                                                                 %
%%%%%%%%%%%%%%%%%%%%%%%%%%%%%%%%%%%%%%%%%%%%%%%%%%%%%%%%%%%%%%%%%%%

\section{Monte Carlo Spectra}
\label{mcspec}

Figure \ref{spectra.basic} shows the emerging photon flux $N(E,\mu)$,
in arbitrary units,
from a Monte Carlo calculation with first
harmonic scattering only and zero natural line width.  The injected
photon flux $N_i(E,\mu)$ is a power law with $s=1$.  Injection is at
the origin of a hot spot with radius $r_h=0.1\ R_*$.  At the stellar
surface the magnetic field strength is $B_{o,12}=1.7$ and the electron
density is $n_{eo}=10^{17}\ {\rm cm^{-3}}$.  The calculation uses the
approximate velocity profile of eq. (\ref{bfapprox2}).  For
comparison, we also plot a pure absorption spectrum,

\begin{equation}
N_{abs}(E,\mu) = N_i(E,\mu)\ \exp \left [-\sum_n {\tau_n(E,\mu)} \right ]\ ,
\label{nabs}
\end{equation}

\noindent
where in this case the sum stops at $n=1$.

\medskip
The motion of the plasma has a significant impact on the
spectrum.  The line centers are shifted from their rest frame values
$E \approx nE_B$, according to
eq. (\ref{eblab}).  In
addition, the high plasma velocity beams scattered photons
along the magnetic field.  Consequently,
when viewed perpendicular to the field [panel (a)], the scattered
spectrum is nearly identical to the absorption spectrum, since photons
absorbed at large angles to the field are re-emitted at smaller ones.
As $\mu$ approaches $\beta_{Fo}=1/2$, the cutoff energy $E_c$ is near
the line center and plays an important role in determining the
properties of the line, as we note in Section \ref{skine}.  Since $E_c$
falls with altitude, photons with $E<E_c$ can escape the line core by
diffusing upwards until $E>E_c$.  The photons then escape the
line-forming region without further scattering.  The escape of photons
in the blue wing is responsible for the spike in panel (b).  This
feature becomes broader with increasing $\mu$, due to the enhanced
Doppler broadening of the line when viewed along the field, and to the
large number of scattered photons that emerge at high $\mu$.  As $\mu$
approaches unity, the line is almost entirely filled in by scattered
photons, forming a broad emission-like feature.

\medskip
The features in Figure \ref{spectra.basic} are narrow.  The equivalent
width of the line in panel (a) is $W_{E1}=5.3$ keV, which is
comparable to the first harmonic line width in the {\it Ginga} observations
of GB880205, $W_{E1}=3.7$ keV.  The reason the line is so narrow, in
spite of the variation of the magnetic field with altitude, is that
the vast majority of scatters occurs in a region close to the surface
where the variation of the field is small.  We illustrate this in
Figure \ref{nscat}, which shows the number of scatters as a function
of altitude $dN_s/dz$, in arbitrary units.  We plot $dN_s/dz$ for the
simulation in Figure \ref{spectra.basic}, as well as for simulations
with first three harmonic and continuum scattering and both
$\beta_F=0.5\ (1+\mu_o)$ and a self-consistent velocity profile.  Adding
the higher harmonics increases the number of scatters.  In all cases,
few scatters occur above $\tilde{z}=1$.  The decrease in the number of
scatters with altitude is due to the escape of photons through the
sides of the cylindrical line-forming region.  The decrease is
analogous to the decrease in the number of 
of {\it unscattered} photons, described by eq. (\ref{angledist}).  

\medskip
Figure \ref{spectra.fnlw} shows that even though the natural line
width is small, it has a noticeable impact on spectra formed in an
outflow, and needs to be included in radiation transfer calculations.
The figure shows the emerging photon flux
from simulations with a 1/E spectrum injected at the origin of a hot
spot with radius $r_h=0.1\ R_*$.  At the surface the magnetic field
strength is $B_{o,12}=1.7$ and the electron density is
$n_{eo}=10^{17}\ {\rm cm^{-3}}$.  The simulations use 
self-consistent velocity profiles.  When the finite natural line width
is included, the scattering profile is no longer zero above $E_c$.
Consequently, photons in the blue wing spike at $\mu \approx 1/2$ do not
escape the line-forming region immediately.  They continue to scatter
to larger values of $\mu$, depopulating the spike and filling in
the lines at higher $\mu$ to a greater extent than when $\Gamma_r=0$.

\medskip
When the photons are injected uniformly across the hot spot instead of
at the origin, the lines become shallower.  This is due to the
reduction in the optical depth seen by photons that are injected
closer to the edges of the hot spot.  We illustrate this in Figure
\ref{spectra.disk} for a line-forming region with the same parameters
as Figure \ref{spectra.fnlw}.

\medskip
Adding continuum and second and third harmonic scattering to the
simulation of Figure \ref{spectra.disk} yields the spectra shown in
Figure \ref{spectra.hiharm}.  For comparison, we also plot the corresponding
spectra for
first harmonic scattering only and pure absorption.  The higher harmonic
features in the scattering spectrum are almost identical to the
absorption features.  This is because the scattering cross sections
favor Raman scattering at the higher harmonics; i.e. most photons
absorbed at $n$=2 and 3 are re-emitted as two or three spawned photons at
lower harmonics (\citeNP{bs82}).  Consequently the scattering profiles
suffice to explain the properties of the higher harmonic lines.  As $\mu$
increases the lines become wider due to Doppler broadening, but shallower
due to the $(1-\mu^{r2})^{n-1}$ dependence of the scattering cross
section.  The asymmetric shape is characteristic of lines formed in an
outflow --- as the altitude increases, both the cyclotron energy and
the plasma density decrease, resulting in an extended tail in the red
wing.  This 
is apparent, for example, in the second harmonic line in panel (b).
At the first harmonic, the
absorption-like features at small $\mu$ are shallower and the
emission-like features at large $\mu$ are more sharply peaked than in
the simulations with first harmonic scattering only.  This is because
additional photons are present at the first harmonic due to
spawning.  These photons fill in the absorption-like features and
build up the emission-like ones.

\medskip
In Figure \ref{spectra.scon} we compare the spectrum of Figure
\ref{spectra.hiharm}, calculated from a self-consistent velocity
profile, to a spectrum based on the approximate velocity profile of
eq. (\ref{bfapprox2}).  The flow velocity at the surface is
$\beta_F=0.26$ in the self-consistent calculation, compared with 0.50
in the approximation.  Because of the smaller plasma velocity near the
surface, the line centers are shifted closer to their rest frame
values.  In addition, the lines are considerably narrower in the
self-consistent calculation.  For $0<\mu<1/8$ [panel (a)], $W_{E1}=3.7$
keV in the self-consistent calculation, compared to 5.1 keV in the
approximation.  The reason for the narrower line is clear from
eq. (\ref{nez}), which shows that the
plasma density $n_e(z)$ is proportional to the ratio
$\beta_{Fo}/\beta_F(z)$.  Figure \ref{bf_vs_z.goh} shows that
this ratio is much smaller in self-consistent calculations.
Consequently, for a given plasma density at $z=0$, the plasma density
at $z>0$ is smaller in the self-consistent calculation, as is the
optical depth. 

\medskip
Figure \ref{spectra.temp} compares the self-consistent calculation of
Figure \ref{spectra.scon}, which has temperature $k_B T_e=0.25\ E_B$, to a
calculation with $k_B T_e=0.05\ E_B$.  As we discuss in Sections
\ref{sPdistf} and \ref{sprofiles}, the effect of the temperature on
the spectra is negligible for the parameters used here.

\medskip
Figure \ref{spectra.edens} illustrates the effect of the surface
electron density on the spectra.  The figure shows the photon flux
emerging from a line-forming region with $r_h =0.05\ R_*$ and
$B_{o,12}$=1.7.  The continuum spectrum of GB880205 is injected
uniformly across the hot spot.  The calculations include scattering in
the continuum and the first three harmonics, finite natural line
width, and a self-consistent velocity profile.  The surface electron
densities are $n_{eo}= 10^{16},\ 10^{17}$, and $10^{18}\ {\rm
cm^{-3}}$.  As expected, both the full-width half-maxima and $|W_E|$
increase with the plasma density.  However, even at $n_{eo}=10^{18}\
{\rm cm^{-3}}$, the equivalent widths are modest, with $W_{E1}=5.2$
and $W_{E2}=7.9$ keV for $0<\mu<1/8$.

\medskip
We plot spectra from self-consistent calculations with {\it polarized}
cross sections in Figure \ref{spectra.pol}.  The line-forming region
has $r_h=0.1\ R_*$, $B_{o,12}=1.7$, and $n_{eo}=10^{17}\ {\rm
cm^{-3}}$.  The calculations include continuum and first three
harmonic scattering with finite natural line width.  The photons,
injected uniformly across the hot spot and with the continuum spectrum
of  GB880205, have 
initial polarizations $P_l=-1$, 0, and 1.  The photons at the first
harmonic scatter and switch polarization modes many times before
escaping.  Consequently, the properties of the first harmonic lines
are not affected very much by the use of polarized cross sections or
the choice of $P_l$.  The scattering cross sections
[eq. (\ref{sigrest})] explain the properties of the higher harmonic
features.  Because of the $\mu^{r2}$ dependence of the cross section
for parallel mode photons, the second and third harmonic lines are
very shallow near $\mu^r=0\ $ (i.e. $\mu=\beta_{Fo} \approx 0.3$) when
$P_l=-1$.  The depth 
of the higher harmonic lines increases with the fraction of
perpendicular mode photons. 

%%%%%%%%%%%%%%%%%%%%%%%%%%%%%%%%%%%%%%%%%%%%%%%%%%%%%%%%%%%%%%%%%%%
%                                                                 %
%      Discussion                                                 %
%                                                                 %
%%%%%%%%%%%%%%%%%%%%%%%%%%%%%%%%%%%%%%%%%%%%%%%%%%%%%%%%%%%%%%%%%%%

\section{Discussion}
\label{discuss}

Fenimore et al. (1988\nocite{fenimore88}) report that the two
harmonically spaced lines observed by {\it Ginga} in GB880205 have
equivalent widths $W_{E1}=3.7$ keV and $W_{E2}=9.1$ keV.  As we show in
Section \ref{mcspec}, cyclotron scattering in an outflow can easily
create lines with comparable equivalent widths, due to the escape of
photons through the sides of the cylindrical line-forming region.  For
example, the 
first two harmonic lines in the spectrum in Figure
\ref{spectra.edens}b with $n_{eo}=10^{18}\ {\rm cm^{-3}}$ and $3/8 <
\mu < 1/2$ have equivalent widths $W_{E1}=2.3$ keV and $W_{E2}=8.8$
keV.  Clearly, the interpretation of the observed features as
cyclotron lines does not rule out burst sources in the galactic
corona, provided the photon source is a small fraction of the stellar
surface.  Thus, the 
cyclotron line interpretation is consistent with the BATSE brightness 
and sky distributions, which suggest that if bursters are galactic, they
are in the corona.  We are currently
fitting the outflow model to the {\it Ginga} data in greater detail by
folding the model spectra through the detector response matrices and
calculating $\chi^2$ (see \citeNP{fenimore88}; \citeNP{wang89};
\citeNP{briggs96a}). 

\medskip
In addition to generating cyclotron lines comparable to the
absorption-like features observed by {\it Ginga}, the outflow model
predicts the formation of emission-like features.  Some static models
predict double-peaked emission-like features, consisting of line
shoulders on each side of an absorption-like line at certain viewing
angles when the photon source is embedded in a plasma which is
optically thin in the continuum (\citeNP{freeman92}; \citeNP{ah96};
\citeNP{ilw97}).  But these are quite distinct from the {\it
single}-peaked features that the outflow model predicts.  Mazets et
al. (1981, 1982\nocite{mazets81}\nocite{mazets82}) report observations
of such features by the Konus instruments.  The line candidates
observed by BATSE (\citeNP{briggs96b}) also include emission-like
features.  These observations should be compared to the model as soon
as the observed spectra and detector response matrices become
available.

\medskip
All the simulations in the present work use plasma densities that are
small enough so that the line-forming region is optically
thin in the continuum, i.e. $n_e \lesssim \sigma_T^{-1} r_h^{-1} = 1.5 \times
10^{19}\ {\rm cm^{-3}}\ r_{h,5}^{-1}$.  But for the physical conditions
required for galactic corona models, the plasma
density could be much larger due to the production of
electron-positron pairs.   
The cross section for the two photon pair
production process $\gamma \gamma \rightarrow e^+e^-$ is $\sim
\sigma_T$, corresponding to a pair production rate

\begin{equation}
R_{\gamma\gamma} \sim {\sigma_T L \over
\pi r_h^2c\ \langle E \rangle} = 440\ {\rm cm^{-1}}\ 
L_{42}\ r_{h,5}^{-2}\ \left ({\langle E \rangle \over 1\ {\rm
MeV}} \right )^{-1}\ ,
\label{rgg}
\end{equation}

\noindent
where $L_{42}\equiv L/10^{42}\ {\rm erg\ s^{-1}}$ is the total (x-ray plus
gamma-ray) burst luminosity, and $\langle E 
\rangle$ is the mean photon energy.  In a strong magnetic field,
one photon pair production ($\gamma \rightarrow e^+e^-$) is
also significant.  The polarization averaged production rate for this
process is 

\begin{eqnarray}
R_{\gamma B} & \approx & 0.26\ {\alpha \over \lambda_C}\ b \sin
\theta\ 
\exp \left [- {8 \over 3}\ {m_e c^2 \over E\ b \sin \theta} \right ]
\nonumber \\
& = &
1.1 \times 10^6\ {\rm cm^{-1}}\ B_{12} \sin \theta\ \exp \left [-6.1\ \left(E
\over 10\ {\rm MeV} \right )^{-1} (B_{12} \sin \theta)^{-1} \right ]
\label{rgblo}
\end{eqnarray}

\noindent for $E b \sin \theta/(m_e c^2) \ll 1$, and

\begin{eqnarray}
R_{\gamma B} & \approx & {5 \over 12} \ {\alpha \over \lambda_C}\ \sin
\theta\ \left ( {E \over m_e c^2}\ b \sin \theta \right )^{-1/3}
\nonumber \\
& = & 2.2
\times 10^7\ {\rm cm^{-1}}\ \sin \theta \left( {E \over 1\ {\rm GeV}}
\right )^{-1/3}\ (B_{12}\sin \theta)^{-1/3}\ 
\label{rgbhi}
\end{eqnarray}

\noindent for $E b \sin \theta/(m_e c^2) \gg 1$, where $\lambda_C
\equiv \hbar / (m_e c)$ is the Compton wavelength (\citeNP{te74};
\citeNP{dh83}; \citeNP{bh84}; \citeNP{meszaros92}).  Clearly the optical
depths for both the one and two photon processes are much larger than
unity for the magnetic fields and photons densities of interest in
galactic corona gamma-ray burst models (\citeNP{schmidt78};
\citeNP{bl87}).  In addition to providing large plasma densities, pair
production could truncate the spectrum at the pair production
threshold, 1 MeV.  Truncation is not consistent with observations by
BATSE (\citeNP{band93}), COMPTEL (\citeNP{winkler93}), and EGRET
(\citeNP{schneid92}; \citeNP{kwok93}; \citeNP{sommer94}) of photon energies
up to 1 GeV with no evidence of a spectral cutoff or rollover at
$\approx$ 1 MeV.

\medskip
The properties of lines formed in an outflow that is optically thick
in the continuum are not very well known.  The full-width half-maxima
of the lines increase with optical depth like $(\ln \tau)^{1/2}$.
This line broadening, though logarithmic, is significant if the plasma
density increases by a factor $\gtrsim 100$; it might be sufficient to
prevent the formation of narrow lines.  However, parallel mode photons
with $\mu \approx \beta_F$ escape the line-forming region easily since
their resonant scattering cross section is $\propto \mu^{r2}$.
Therefore, the spectrum may possess narrow cyclotron features when
viewed along this line of sight.  The photon flux along this line of
sight is $\sim$ one-tenth the flux at $\mu=1$ because nearly all
photons scatter at these optical depths and are beamed along the
field.  Consequently, if the line of sight is at a large angle to the
field, the observed flux is reduced and narrow lines are present.  If
the line of sight is at a small angle to the field, the observed flux
is large.  The absence of narrow features in gamma-ray burst spectra
calculated from the ``CUSP'' model, in which the spectrum is formed by
the {\it C}yclotron {\it U}p {\it S}cattering {\it P}rocess
(\citeNP{vd91}; \citeNP{cl92}) suggests that no narrow cyclotron feature is
visible when the line of sight is at a small angle to the field.  In
short, if the cyclotron lines are formed in an outflow that is
optically thick in the continuum, the presence of narrow lines is inversely
correlated with the brightness of the burst.  This correlation may be
masked by selection effects: lines are easier to detect in brighter
bursts.  Whether spectra formed under these circumstances are
consistent with observations is an open question.  Additional
simulations are required, using cross sections that are appropriate
for plasmas that are optically thick in the line wings.

\medskip
The plasma densities in corona models may be smaller than
eqs. (\ref{rgg})--({\ref{rgbhi}) suggest
(for a review see \citeNP{hl90}).  One often cited
possibility is that the entire photon spectrum is produced at the
magnetic polar cap, but the photons are beamed into a small
angle.  Another is that the x-rays and gamma-rays are produced in two separate
physical regions, by two different mechanisms.

\medskip
If a plasma with bulk motion corresponding to $\gamma_F$
produces the spectrum, then the
photons are beamed into an angle $\sim \gamma_F^{-1}$.  The
kinematics of two photon pair production requires $E_1 E_2 (1-\cos
\theta_{12}) \geq 2 m_e^2 c^4$, where $E_1$ and $E_2$ are the energies
of the two photons and $\theta_{12}$ is the angle between them.  
A beaming angle $\sim 0.1$ (i.e. $\Gamma_F \sim 10$)
raises the pair production  
threshold enough to be consistent with the observed spectra above 1
MeV (\citeNP{schmidt78}; \citeNP{bh93}; \citeNP{hb94}; \citeNP{harding94}).
Unfortunately, it is unlikely that the observed cyclotron features are
produced in a plasma with such a large Lorentz factor.  According to
eq. (\ref{eblab}), the creation of a
cyclotron first harmonic at 20 keV when $\gamma_F=10$ and $\theta=0.1$
requires a magnetic field $B_{12} \approx 0.2$, which is about an
order of magnitude smaller than the fields we use in the present work.
Since the ratio of the second and first harmonic scattering cross
sections $\sigma_2/\sigma_1 \propto B$, the ratio $W_{E2}/W_{E1}$ is
much smaller for $B_{12} \approx 0.2$ than in the
{\it Ginga} observations.  To obtain the observed spectrum, the lines
must be formed in a plasma in a separate location from the plasma
responsible for beaming the radiation.  The plasma in the line-forming
region must also be moving more slowly than $\gamma_F=10$.

\medskip
This leads to the second possibility, that the gamma-ray burst
spectrum consists of two 
components produced in two distinct physical regions: a soft ($E
\lesssim 1$ MeV) component produced near the magnetic pole and a hard
component ($E \gtrsim 1$ MeV), probably produced at some distance above the
surface (\citeNP{lamb82}; Katz  
1982, 1994\nocite{katz82}\nocite{katz94}).
If
this is the case, then the photons in the line-forming region have
energies below the pair production threshold and the plasma density
can remain small.   Since the region where the gamma-rays are produced
is larger and has a smaller magnetic field than in models
where the gamma-rays are produced at the polar cap, there
is not sufficient pair production to truncate the spectrum at 1 MeV.
A recent report by Chernenko and Mitrofanov 
(1995\nocite{cm95}) of evidence for two components in the spectrum of
GB881024 is intriguing.  The soft component dominates the 
spectrum for $E < 250$ keV; the hard component is dominant at higher
energies.  However, the question of exactly where and how the high energy
component is produced remains open.
  
\medskip
The cyclotron line interpretation of the observed spectral features
strongly supports a galactic origin for at least some gamma-ray
bursts.  Recent 
reports of fading x-ray and optical counterparts apparently
associated with GB970228 raise hopes that we will soon know the
distance scale to the burst sources.  
If it turns out that some bursts are galactic, then our understanding
of how cyclotron lines are formed at low energies will play an important
role in constraining models of how the spectrum is formed at all
energies.

\acknowledgements
We wish to thank Jeff Benensohn, Tomek Bulik, Peter Freeman, Carlo
Graziani, Cole Miller, Guy Miller, Lucia Mu\~noz-Franco, Jean
Quashnock, Paul Ricker, Jane Wang, Rob Wickham, Tom Witten, and
Shanqun Zhan for invaluable discussions.  We are grateful to Tsvi
Piran and Stan Woosley for pointing out the necessity of considering
the impact of pair production on our model.  We also wish to thank
Corbin Covault, Mark Oreglia, and Jonathan Rosner for their careful
reading of the manuscript and their helpful suggestions.  
This work was supported in part by NASA Grants  NAGW-830, NAGW-1284,
and NAG5-2868.  MI is happy to acknowledge the support of a NASA
Graduate Student Researchers Fellowship under NASA Grant NGT-5-18.
This work partially satisfies the requirements for the degree of
Ph.D. from the University of Chicago for MI.

\begin{figure}
\figcaption{Geometry of the Line-Forming Region.  Photons are injected at
a hot spot with radius $r_h$ located at the magnetic pole of a neutron
star with dipole field $B$.  They scatter with the electrons in the
line-forming region, accelerating them along the field lines.  $\beta$
is the velocity of an individual electron and $\beta_F(z)$ is the bulk
velocity of the flow, where z is the altitude above the stellar
surface.  $\theta$ denotes the angle between the photon direction and
the magnetic field direction.  The angular size of the hot spot is
$\theta_o(z)$ \label{lfr}} 
\end{figure}

\begin{figure}
\figcaption{First harmonic resonant photon energy $E_1$ as a function of
electron momentum $p$ for $B_{12}=1.7$ and several photon orientations
$\mu$.  For a given $\mu$ and $E_1$ the resonant condition has two
roots $p_-$ ({\it solid lines}) and $p_+$ ({\it dotted lines}).  Note
the cutoff energy $E_c$ where $p_-=p_+$ and $\beta=\mu$
(i.e. $\mu^r=0$).  Physical roots do not exist for $E>E_c$.  The
values of $p$ corresponding to $\beta=0$ and $\beta=0.5$ are indicated
by {\it dashed lines}. \label{fecons}}
\end{figure}

\begin{figure}
\figcaption{Radiation-driven acceleration of an electron-proton pair in a
strong magnetic field, according to the model of Mitrofanov and Tsygan
(1982).  The particles are injected at the origin of a hot spot with
radius $r_h$.  Radiation with a 1/E spectrum and X-ray
luminosity $L_x=10^{40}\ {\rm erg\ s^{-1}}$ is injected
semi-isotropically and uniformly across the hot spot.  Left: velocity
$\beta$ as a function of altitude $\tilde{z}$.  Right: Lorentz
factor $\gamma$.  The {\it solid lines} are the $\beta$ and $\gamma$
of a pair with zero initial velocity.  The {\it dotted lines} indicate
the values of $\beta$ and $\gamma$ where the radiation force
$F_{rad}=0$.
\label{b_vs_z}}
\end{figure}

\begin{figure}
\figcaption{Effect of the spectral index $s$ on the velocity profile.
Flow velocity as a function of altitude calculated from the model of
Mitrofanov and Tsygan (1982) for $s$=0.5 ({\it dotted line}), 1.0 ({\it
dashed line}), and 2.0 ({\it dot-dash}) and unpolarized scattering.
The approximate profile $\beta_F=0.5\ (1+\mu_o)$ is shown for
comparison ({\it solid line}).
\label{bf_vs_z.s}} 
\end{figure}

\begin{figure}
\figcaption{Effect of the photon polarization on the velocity profile.
Flow velocity as a function of altitude calculated from the model of
Mitrofanov and Tsygan (1982) for spectral index $s$=1 and polarized
scattering.  The fraction of injected photons in parallel mode is
$P_l=-1.0$ ({\it dashed line}), 0.0 ({\it
dotted line}), and 1.0 ({\it solid line}).
\label{bf_vs_z.pol}} 
\end{figure}

\begin{figure}
\figcaption{Self-consistent Monte Carlo calculations of velocity profiles
for a 1/E spectrum injected into a line-forming 
region with hot spot radius $r_h=0.1\ R_*$, dipole field
$B_{o,12}=1.7$, and surface electron density $n_{e,o}=10^{17}\ {\rm
cm^{-3}}$.  The profile is shown for first harmonic scattering,
point photon source, and natural line width $\Gamma_r$=0 ({\it stars}); first
harmonic scattering, point source, and $\Gamma_r$=finite ({\it
squares}); first harmonic scattering, disk source, and
$\Gamma_r$=finite ({\it dots}); and continuum and first three harmonic
scattering, disk source, and $\Gamma_r$=finite ({\it triangles}).  The
profile calculated from the model of Mitrofanov and Tsygan (1982) is
shown for comparison ({\it solid line}).
\label{bf_vs_z.goh}} 
\end{figure}

\begin{figure}
\figcaption{Effect of electron density on the velocity.  Velocity
profiles for a GB880205 continuum spectrum injected uniformly across a
hot spot with radius $r_h=0.05\ R_*$.  The line-forming region has
dipole field $B_{o,12}=1.7$, electron temperature $k_B T^{\rm f}_e=0.25\
E_B$, a self-consistent velocity profile, unpolarized continuum and
first three harmonic scattering with finite natural line width, and
surface electron densities $n_{e,o}=10^{16}\ {\rm cm^{-3}}$ ({\it
squares}), $n_{e,o}=10^{17}\ {\rm cm^{-3}}$ ({\it dots}), and
$n_{e,o}=10^{18}\ {\rm cm^{-3}}$ ({\it triangles}).  The profile
calculated from the model of Mitrofanov and Tsygan (1982) is shown for
comparison ({\it solid line}).
\label{bf_vs_z.edens}}
\end{figure}

\begin{figure}
\figcaption{Monte Carlo emerging photon number spectra for several
viewing angles for a 1/E continuum spectrum injected at the origin of
a hot spot with radius $r_h=0.1\ R_*$.  The line-forming
region has a dipole field $B_{o,12}=1.7$, electron temperature $k_B
T^{\rm f}_e=0.25\ E_B$, surface electron density $n_{e,o}=10^{17}\ {\rm
cm^{-3}}$, velocity profile $\beta_F=0.5\ (1+\mu_o)$, and unpolarized
first harmonic scattering with zero natural line width ({\it
solid lines}).  Pure absorption spectra ({\it dashed lines}) are shown
for comparison.
\label{spectra.basic}} 
\end{figure}

\begin{figure}
\figcaption{Number of scatters per photon per unit distance $dN_s/dz$, in
arbitrary units, as
a function of altitude above the stellar surface $z/r_h$ for a 1/E
continuum spectrum injected into a line-forming region with dipole
field $B_{o,12}=1.7$, electron temperature $k_B T^{\rm f}_e=0.25\ E_B$,
surface electron density $n_{e,o}=10^{17}\ {\rm cm^{-3}}$, and
unpolarized scattering.  Calculations are shown for continuum and
first three harmonic scattering with self-consistent ({\it solid
line}) and $\beta_F=0.5\ (1+\mu_o)$ ({\it dotted line}) velocity profiles,
and first harmonic scattering with a $\beta_F=0.5\ (1+\mu_o)$ ({\it
dashed line}) velocity profile.
\label{nscat}}
\end{figure}

\begin{figure}
\figcaption{Effect of finite natural line width.  Monte Carlo emerging
photon number spectra for several viewing angles for a 1/E continuum
spectrum injected at the origin of a hot spot with radius
$r_h=0.1\ R_*$.  The line-forming region has dipole
field $B_{o,12}=1.7$, electron temperature $k_B T^{\rm f}_e=0.25\ E_B$,
surface electron density $n_{e,o}=10^{17}\ {\rm cm^{-3}}$, a
self-consistent velocity profile, and unpolarized first harmonic
scattering  with finite ({\it solid lines}) and zero ({\it dotted lines})
natural line width.
\label{spectra.fnlw}} 
\end{figure}

\begin{figure}
\figcaption{Effect of disk source.  Monte Carlo emerging photon number
spectra for several viewing angles for a 1/E continuum spectrum
injected uniformly across a hot spot with radius $r_h=0.1\ R_*$
({\it solid lines}) and at the origin ({\it dotted lines}).  The
line-forming region has dipole field $B_{o,12}=1.7$,
electron temperature $k_B T^{\rm f}_e=0.25\ E_B$, surface electron density
$n_{e,o}=10^{17}\ {\rm cm^{-3}}$, a self-consistent velocity profile,
and unpolarized first harmonic scattering  with finite natural
line width.
\label{spectra.disk}} 
\end{figure}

\begin{figure}
\figcaption{Effect of higher harmonic scattering.  Monte Carlo emerging
photon number spectra for several viewing angles for a 1/E continuum
spectrum injected uniformly across a hot spot with radius
$r_h=0.1\ R_*$.  The line-forming region has dipole
field $B_{o,12}=1.7$, electron temperature $k_B T^{\rm f}_e=0.25\ E_B$,
surface electron density $n_{e,o}=10^{17}\ {\rm cm^{-3}}$, a
self-consistent velocity profile, and unpolarized continuum and first
three harmonic scattering with finite natural line width ({\it
solid lines}).  Spectra for first harmonic scattering  ({\it dotted
lines}) and pure absorption ({\it dashed lines}) are shown for comparison.
\label{spectra.hiharm}} 
\end{figure}

\begin{figure}
\figcaption{Monte Carlo emerging photon number spectra for several
viewing angles for a 1/E continuum spectrum injected uniformly across
a hot spot with $r_h=0.1\ R_*$.  The line-forming region has dipole
field $B_{o,12}=1.7$, electron temperature $k_B T^{\rm f}_e=0.25\ E_B$,
surface electron density $n_{e,o}=10^{17}\ {\rm cm^{-3}}$, unpolarized
continuum and first three harmonic scattering with finite natural line
width, and a self-consistent ({\it solid lines}) and
$\beta_F=0.5\ (1+\mu_o)$ ({\it dotted lines}) velocity profile.
\label{spectra.scon}} 
\end{figure}

\begin{figure}
\figcaption{Effect of electron temperature.  Monte Carlo emerging
photon number spectra for several viewing angles for a 1/E continuum
spectrum injected uniformly across a hot spot with radius
$r_h=0.1\ R_*$.  The line-forming region has dipole
field $B_{o,12}=1.7$, surface electron density $n_{e,o}=10^{17}\ {\rm
cm^{-3}}$, a self-consistent velocity profile, unpolarized continuum
and first three harmonic scattering with finite natural line width,
and electron temperatures $k_B T^{\rm f}_e=0.05\ E_B$ ({\it
solid lines}) and $0.25\ E_B$ ({\it dotted lines}).
\label{spectra.temp}} 
\end{figure}

\begin{figure}
\figcaption{Effect of electron density.  Monte Carlo emerging photon
number spectra for several viewing angles for a GB880205 continuum
spectrum injected uniformly across a hot spot with radius
$r_h=0.05\ R_*$.  The line-forming region has dipole field
$B_{o,12}=1.7$, electron temperature $k_B T^{\rm f}_e=0.25\ E_B$, a
self-consistent velocity profile, unpolarized continuum and first
three harmonic scattering with finite natural line width, and surface
electron densities $n_{e,o}=10^{16}\ {\rm cm^{-3}}$ ({\it solid
lines}), $n_{e,o}=10^{17}\ {\rm cm^{-3}}$ ({\it dotted lines}), and
$n_{e,o}=10^{18}\ {\rm cm^{-3}}$ ({\it dashed lines}).
\label{spectra.edens}} 
\end{figure}

\begin{figure}
\figcaption{Effect of photon polarization.  Monte Carlo emerging photon
number spectra for several viewing angles for a GB880205 continuum spectrum
injected uniformly across a hot spot with radius $r_h=0.1\ R_*$.
The line-forming region has dipole field $B_{o,12}=1.7$,
electron temperature $k_B T^{\rm f}_e=0.25\ E_B$, surface electron density
$n_{e,o}=10^{17}\ {\rm cm^{-3}}$, a self-consistent velocity profile, and
polarized continuum and first three harmonic scattering with finite
natural line width.  The fraction of injected photons in
parallel mode is $P_l=-1.0$ ({\it dashed lines}), 0.0 ({\it dotted lines}),
and 1.0 ({\it solid lines}).
\label{spectra.pol}}
\end{figure}

\begin{deluxetable}{l|c|c|c|}
\tablecaption{Summary of calculations of cyclotron line formation in
radiation-driven outflows associated with gamma-ray
bursts.}
\label{Tcomp}
\startdata
& Miller et al.&
Chernenko \& & 
Present\nl
&1991, 1992
&Mitrofanov 1993 &Work\nl
\tableline
Process& Absorption\ \ \ \ \ & Absorption\ \ \ \ \ & Scattering\ \ \ \ \ \nl\tableline
Continuum Scat.& No& No& Yes\nl\tableline
Natural line width& 0& finite& finite\nl\tableline
Thermal broadening& Yes& No& Yes\nl\tableline
Polarization& No& No& Yes\nl\tableline
Self-consistency& No& No& Yes\nl\tableline
\enddata
\end{deluxetable}

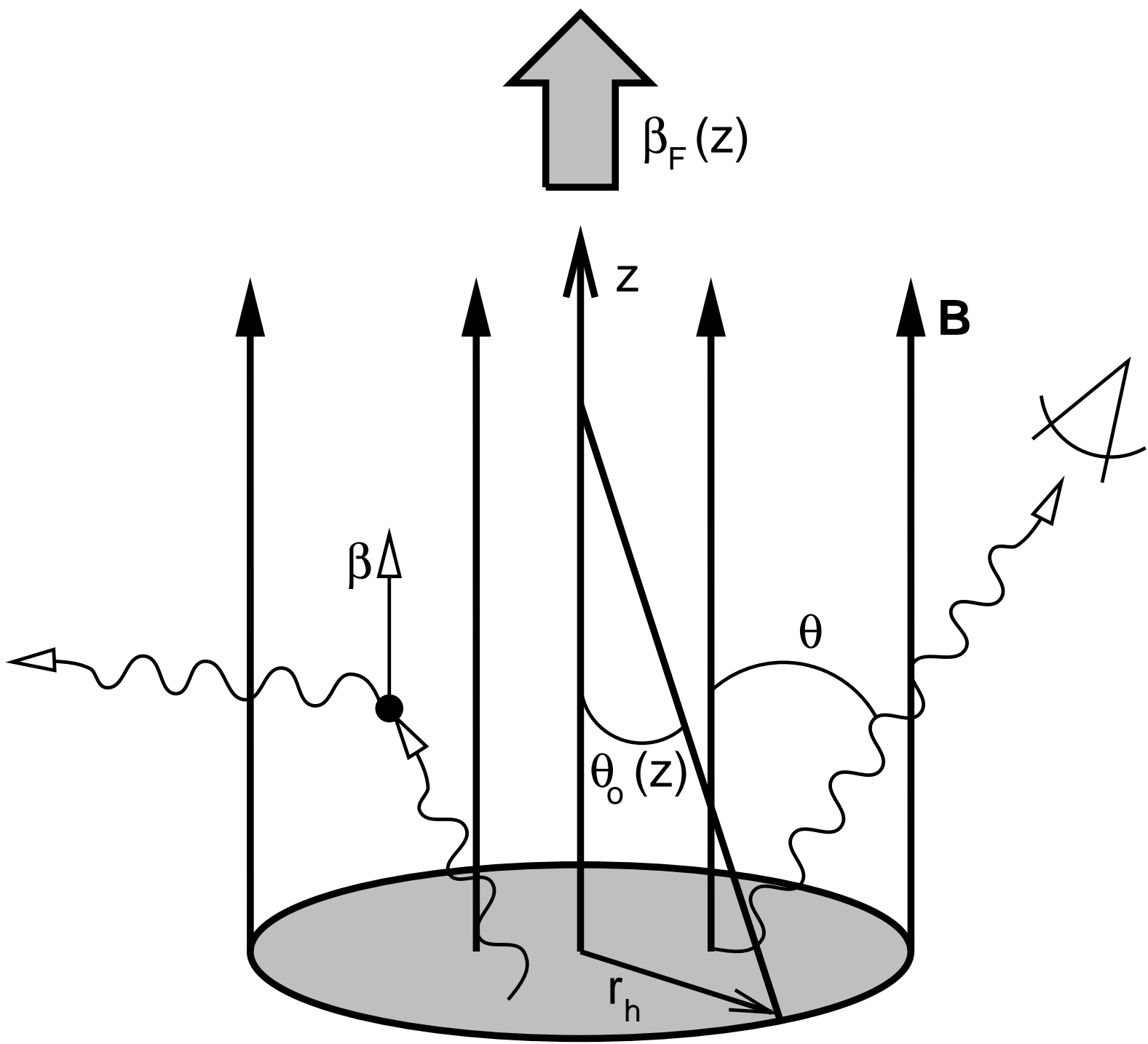
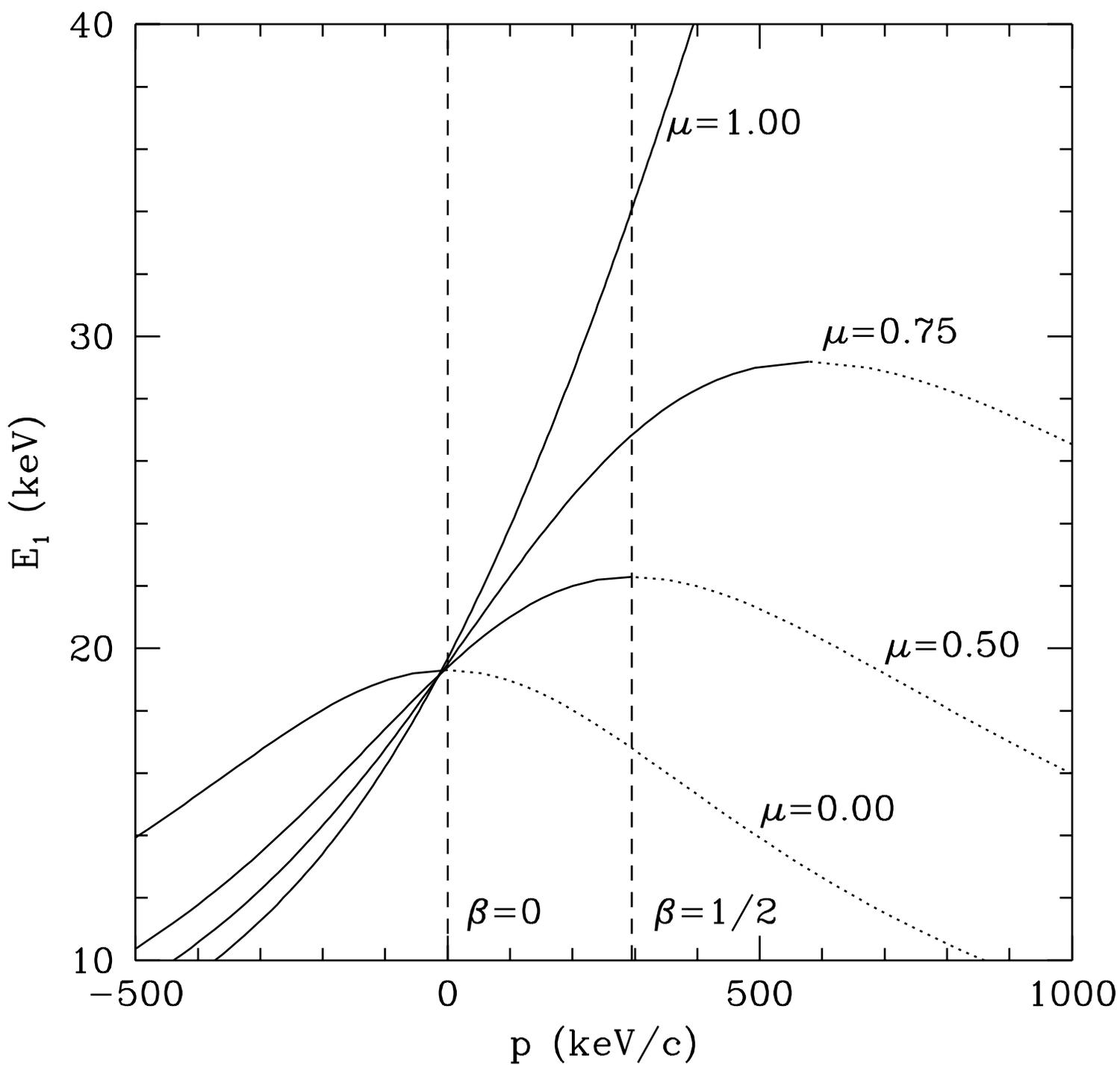
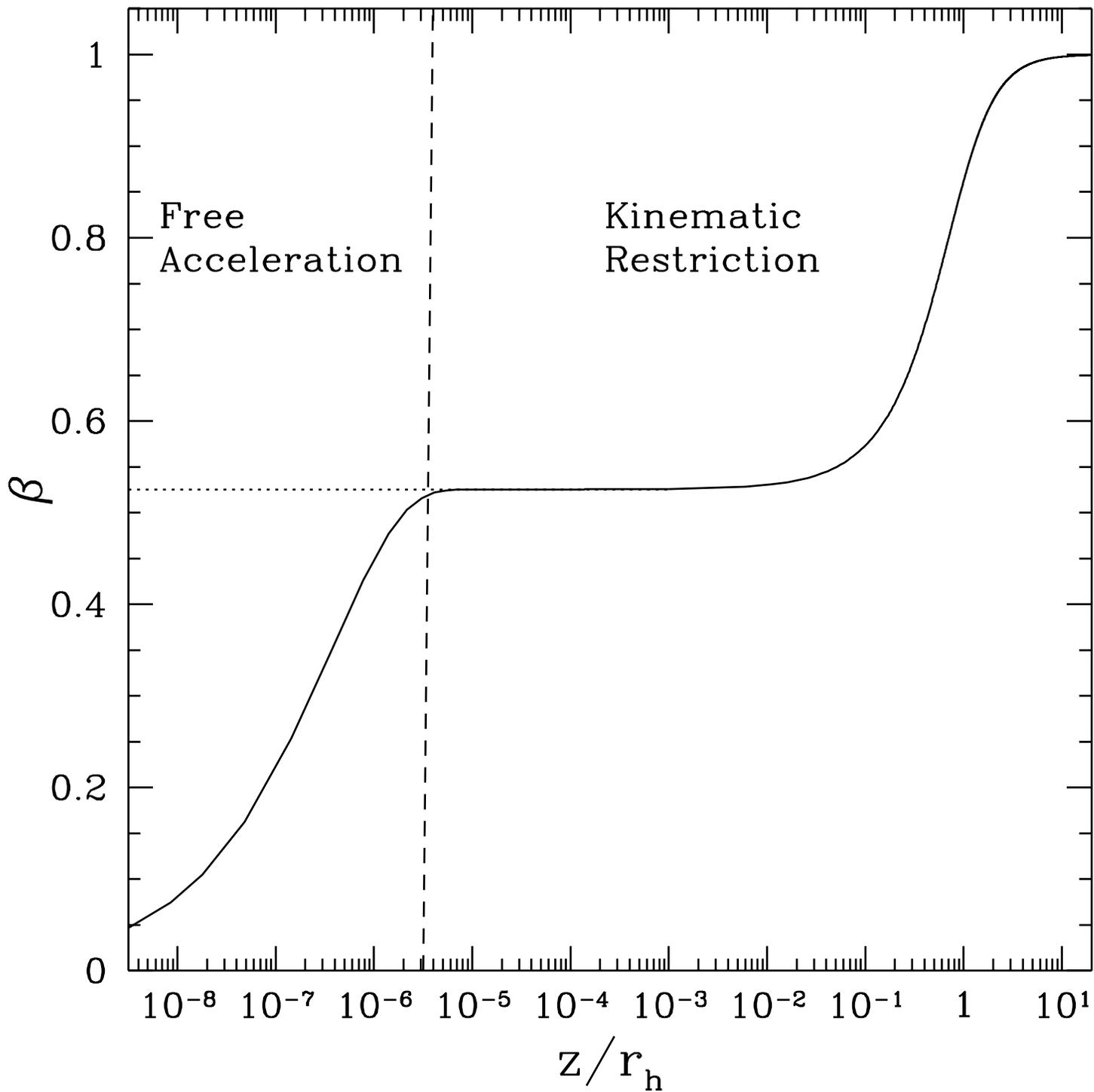
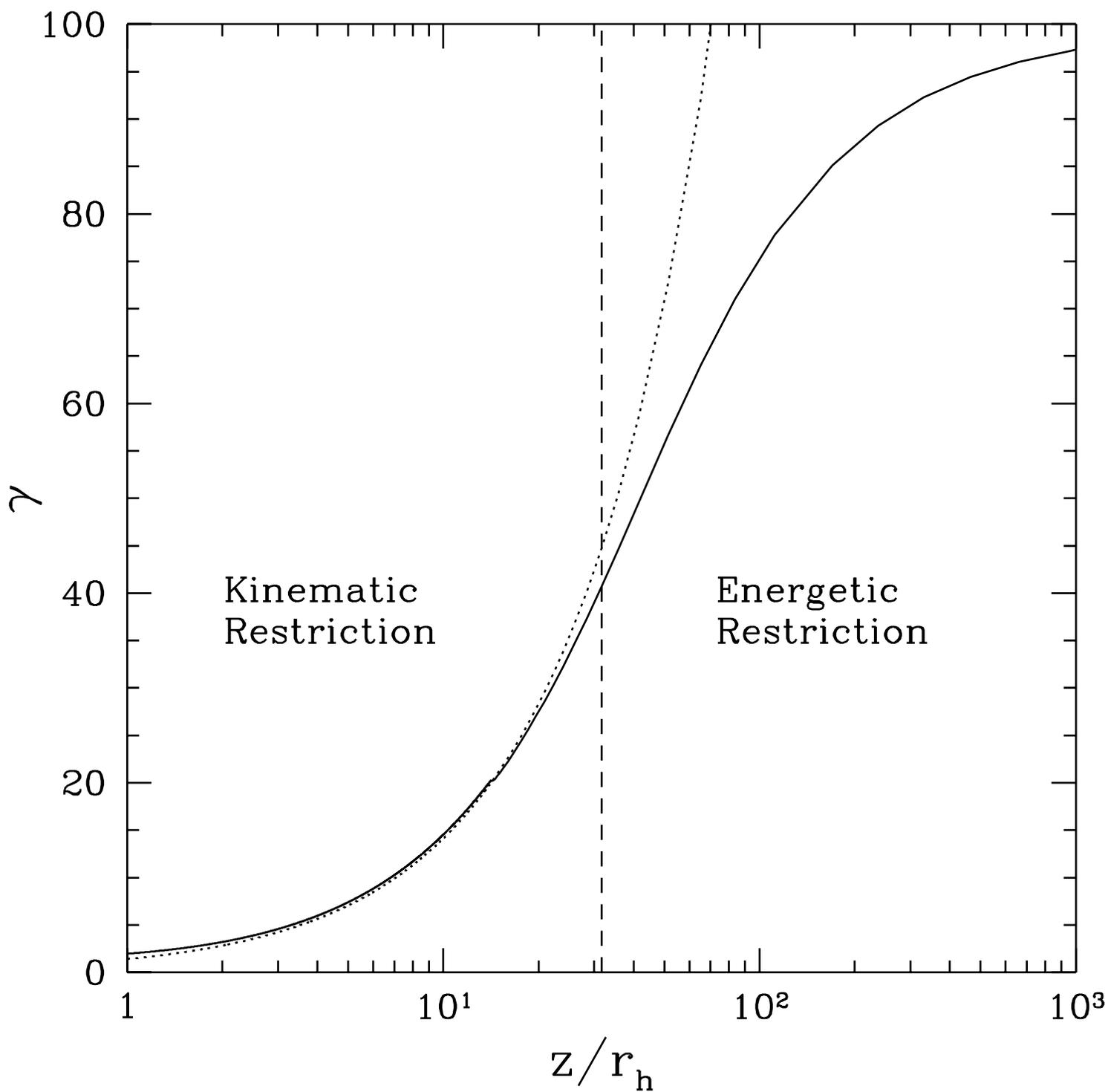
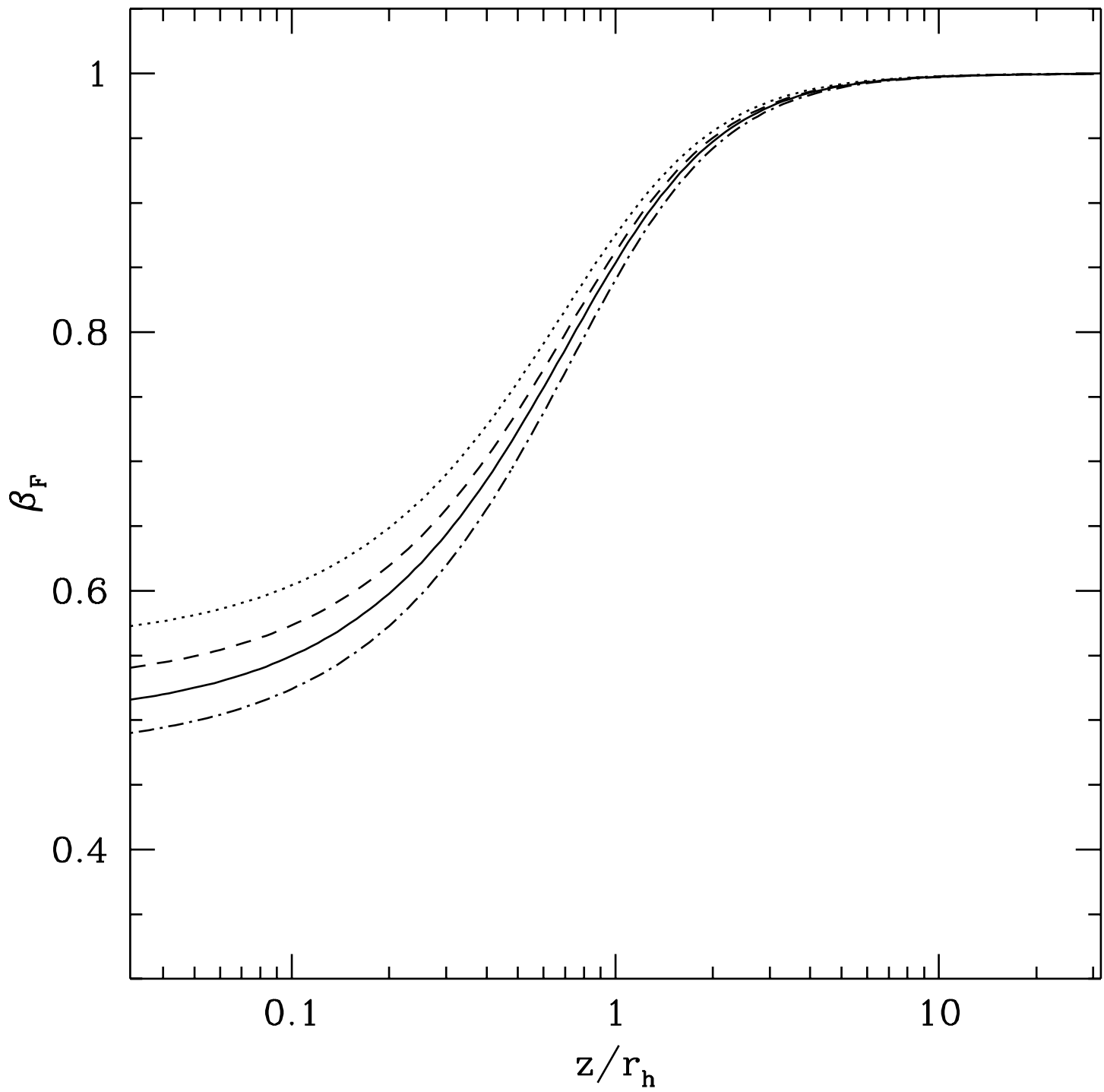
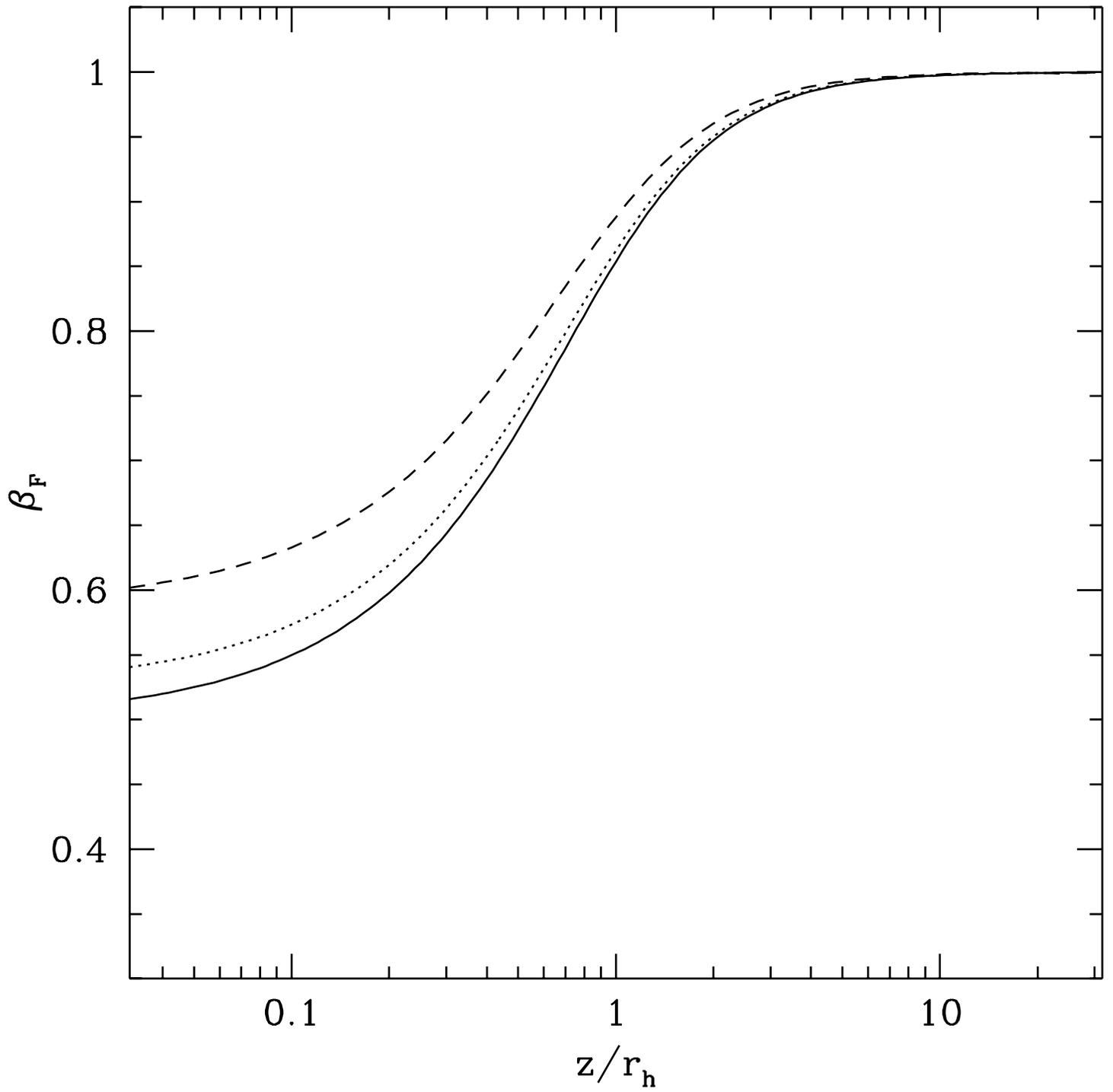
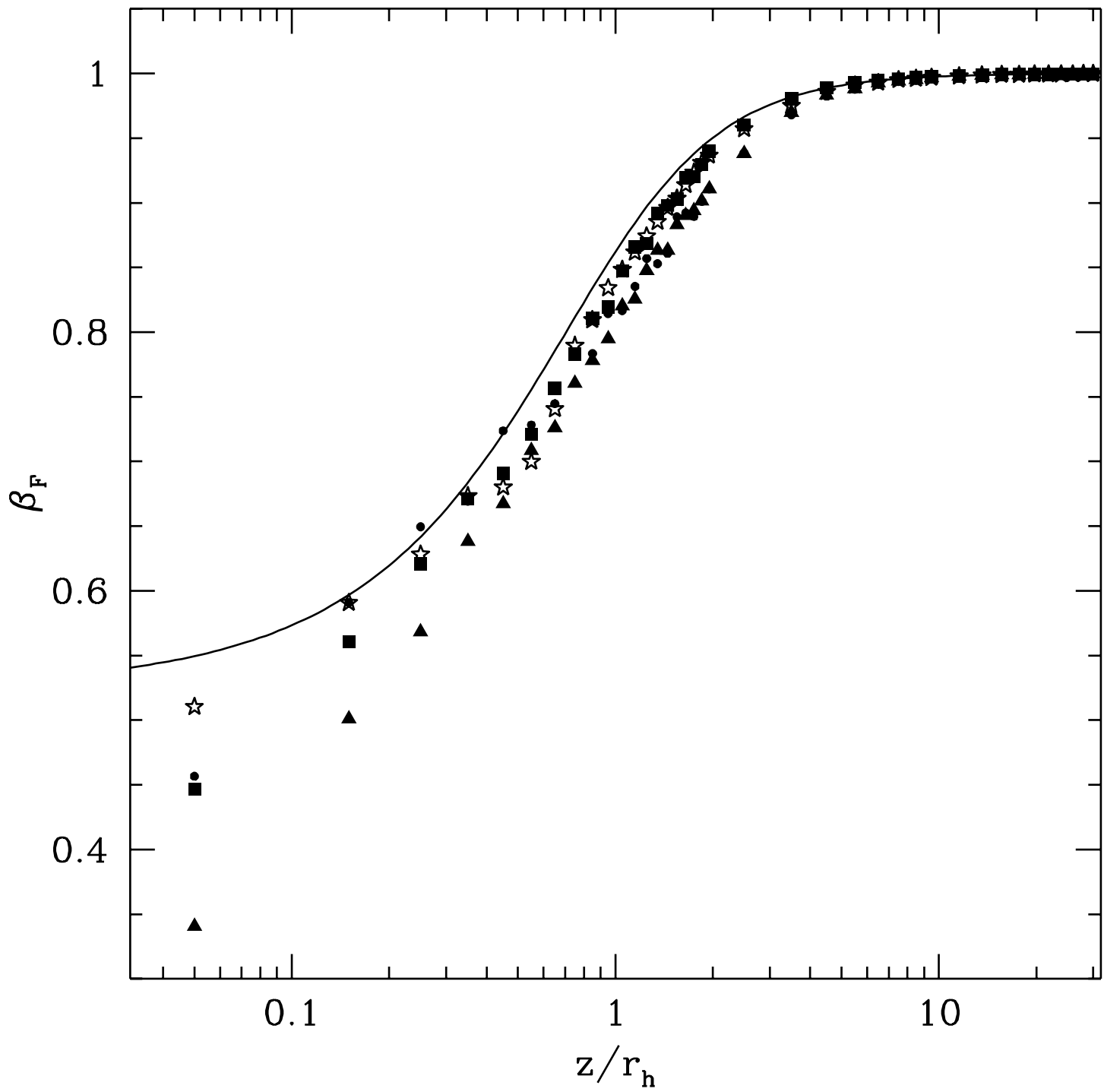
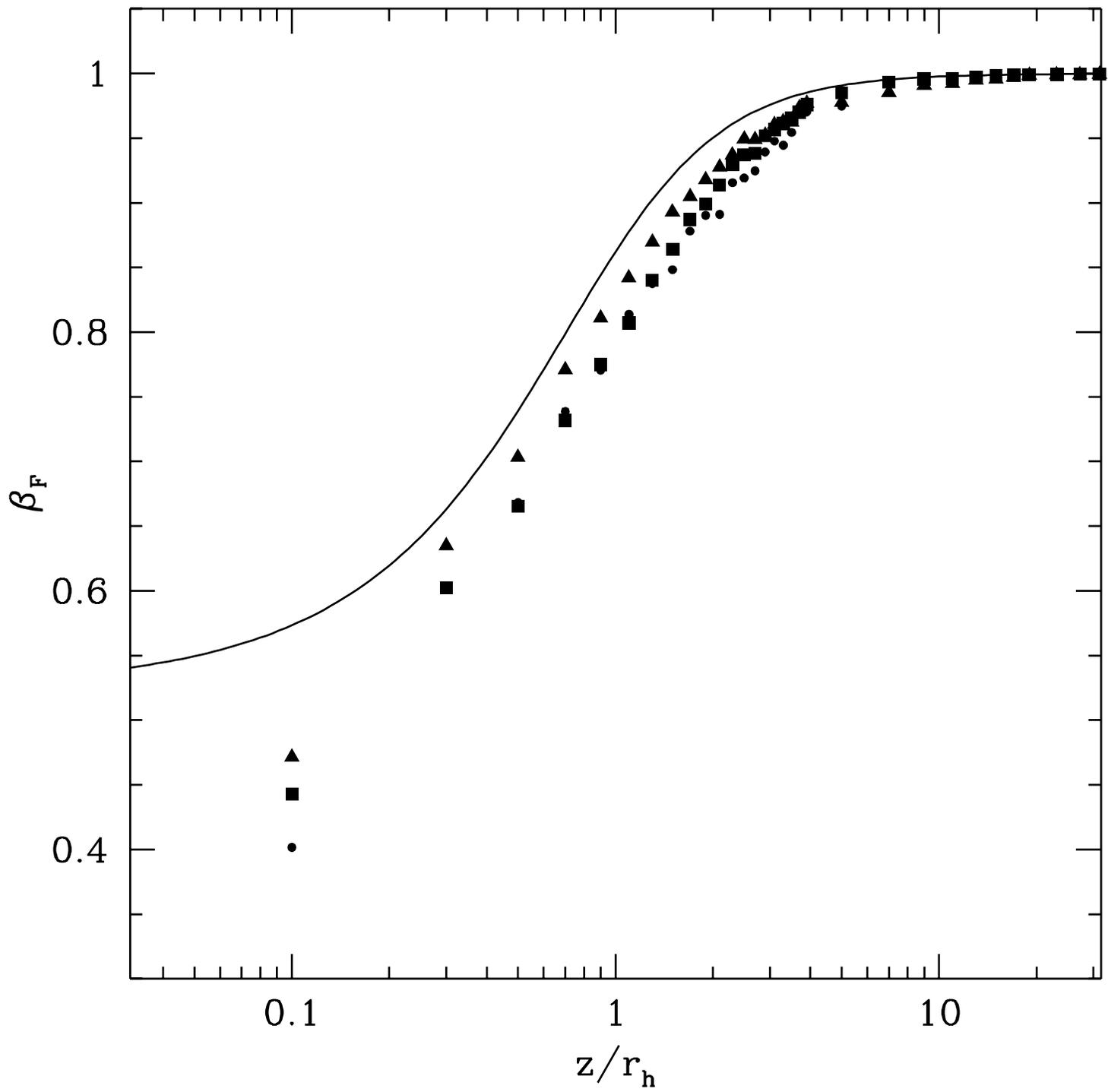
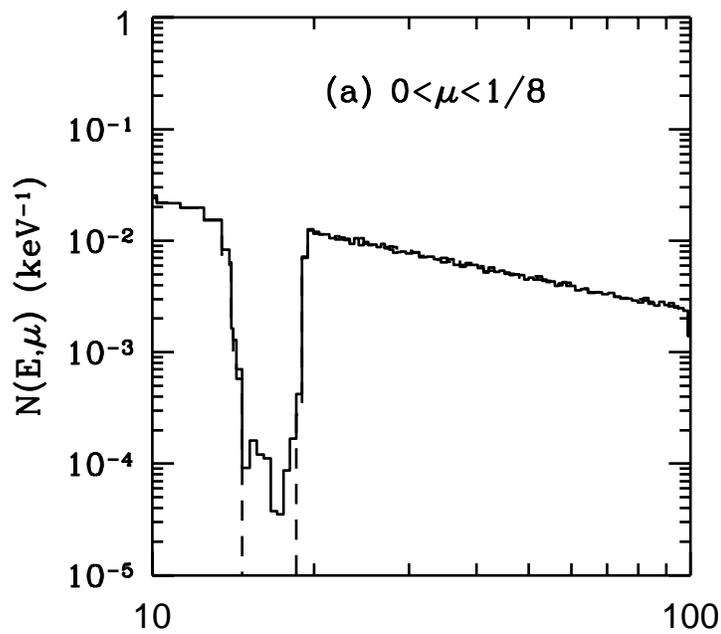
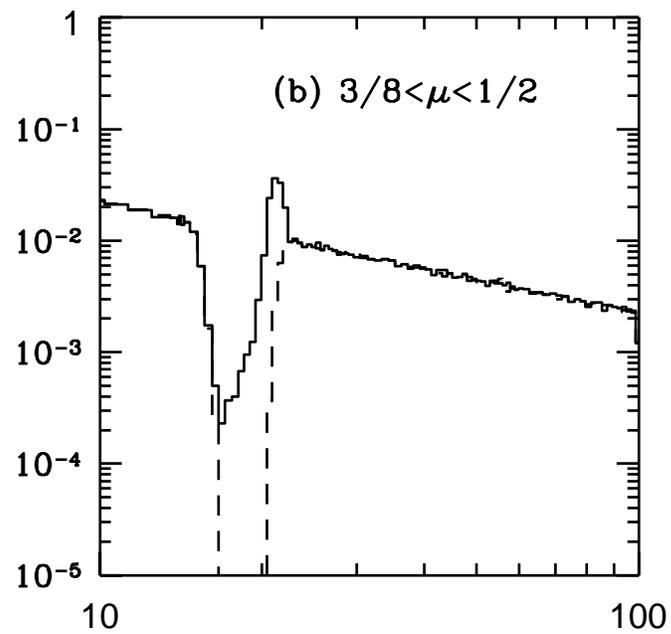
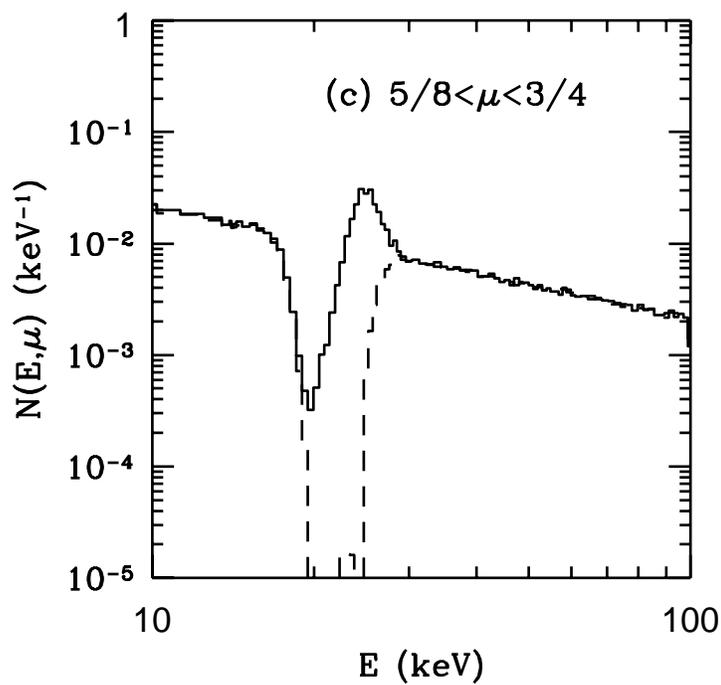
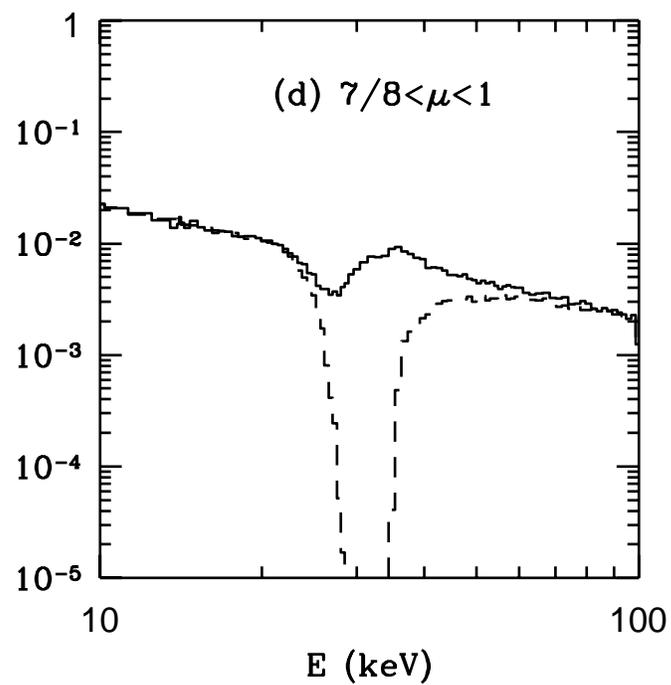
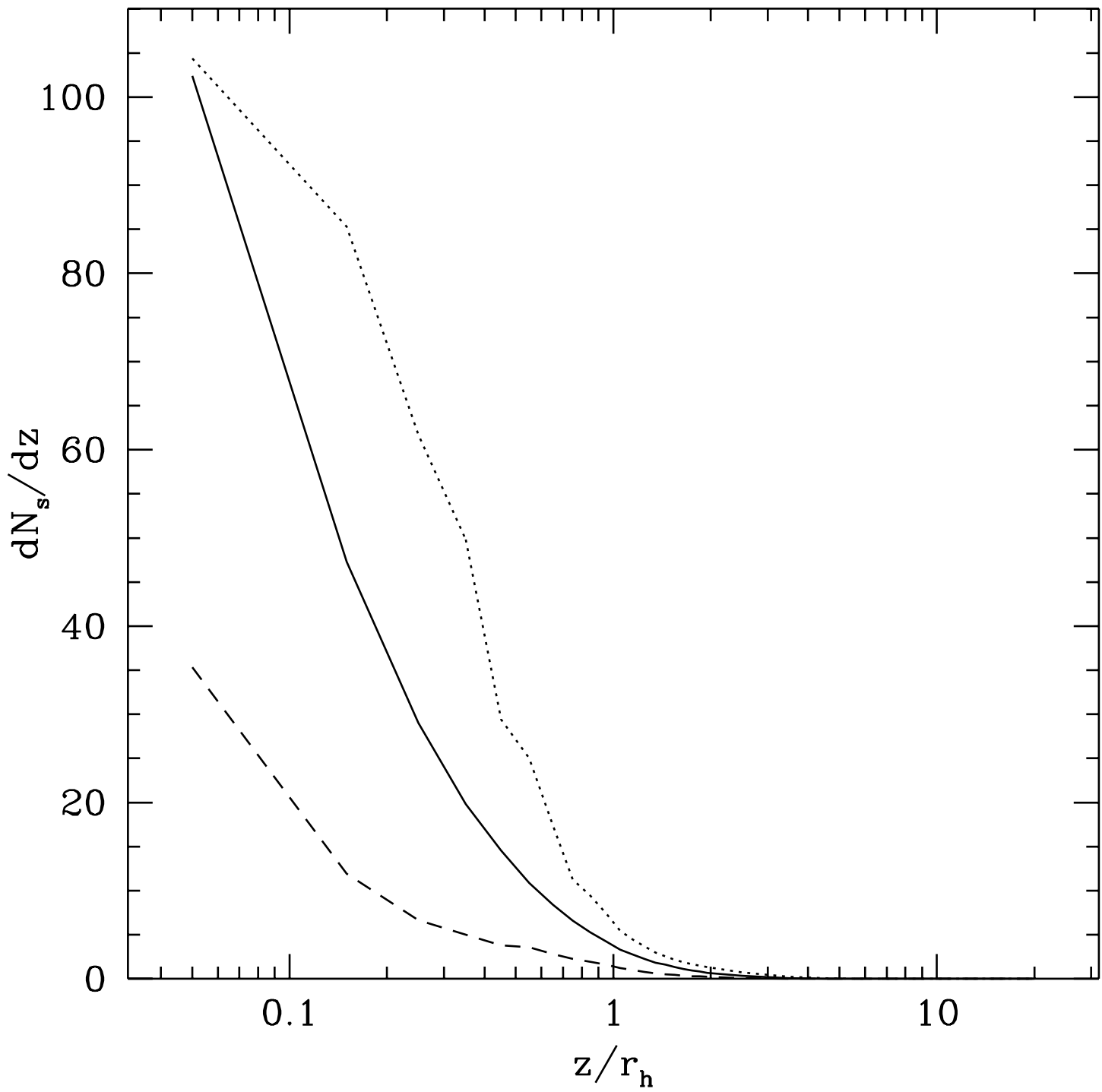
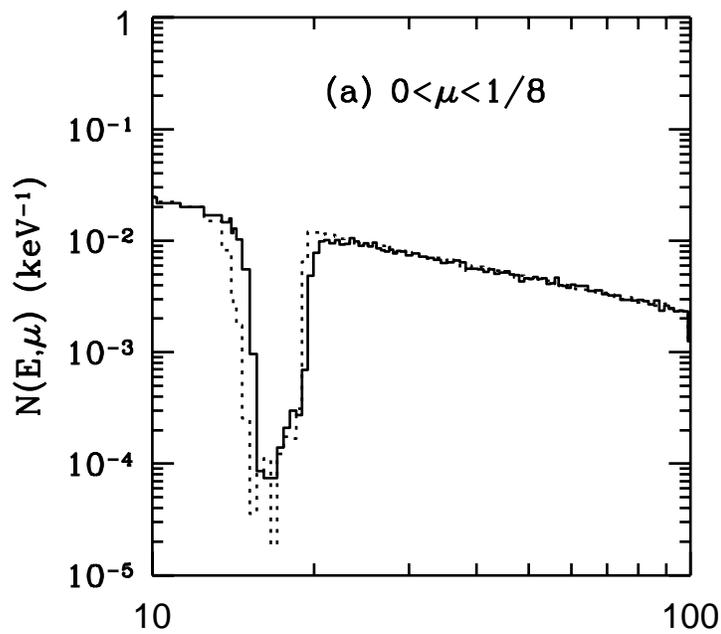
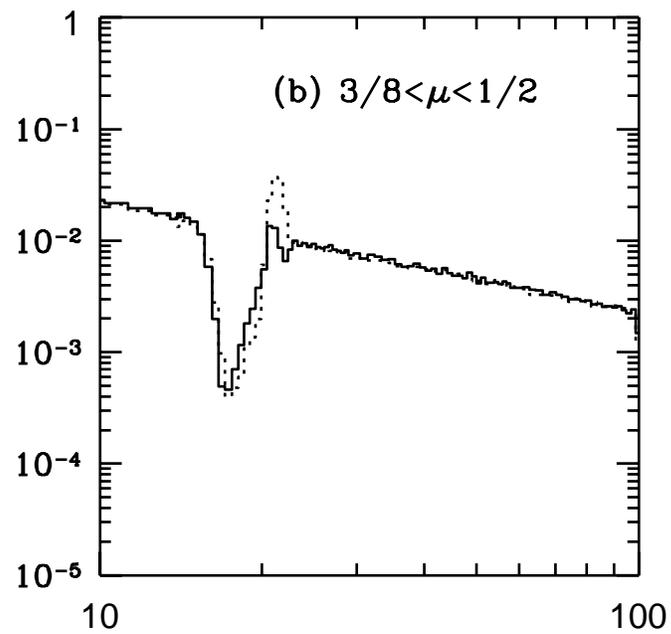
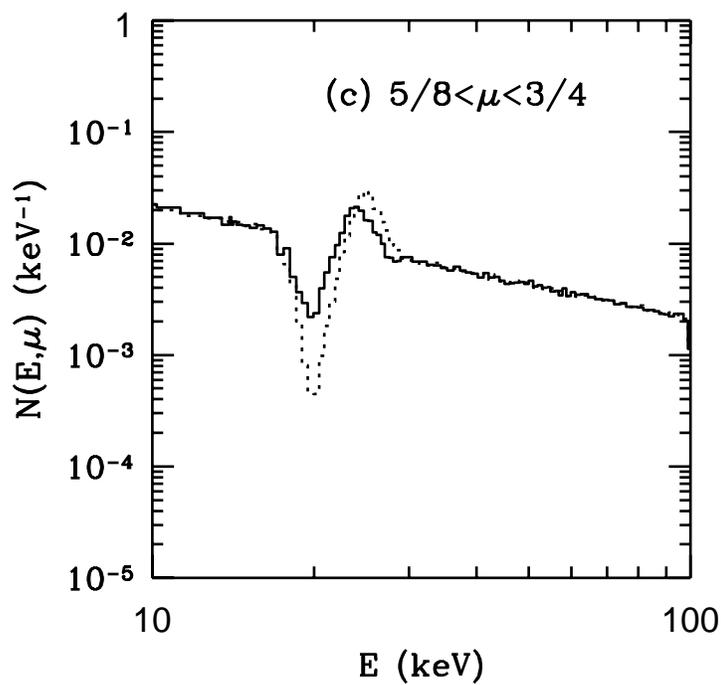
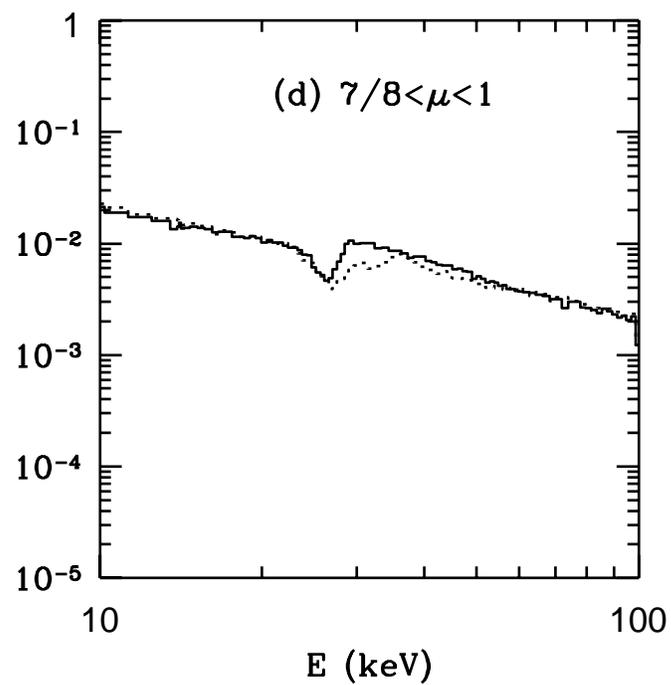
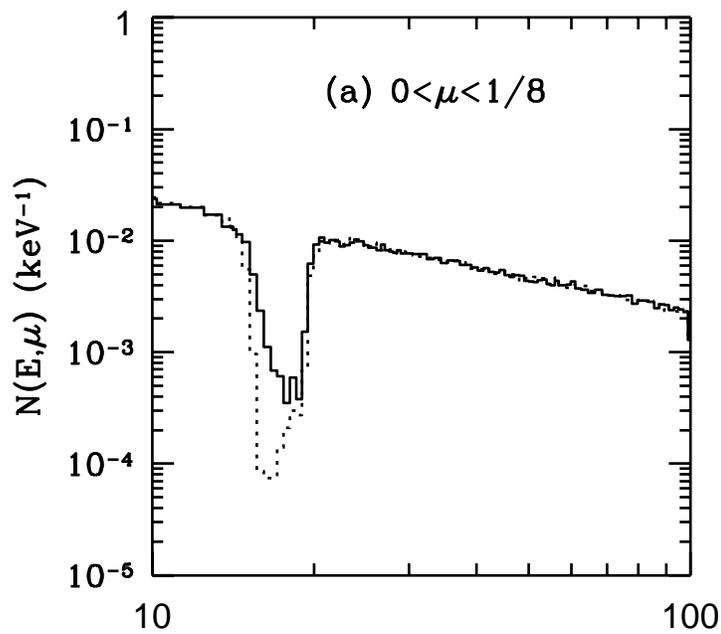
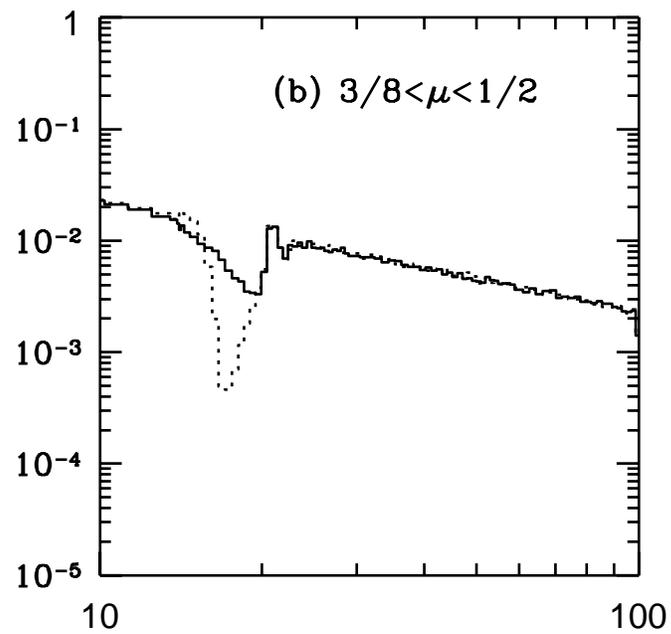
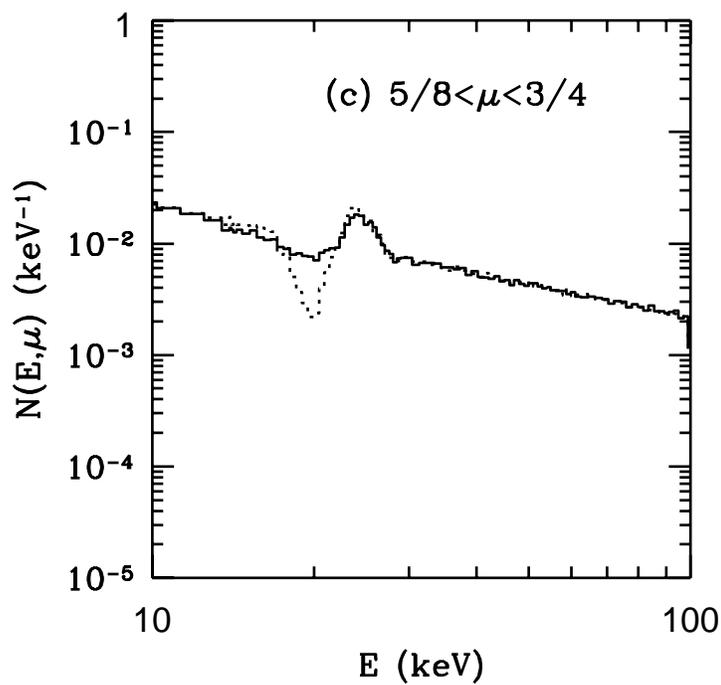
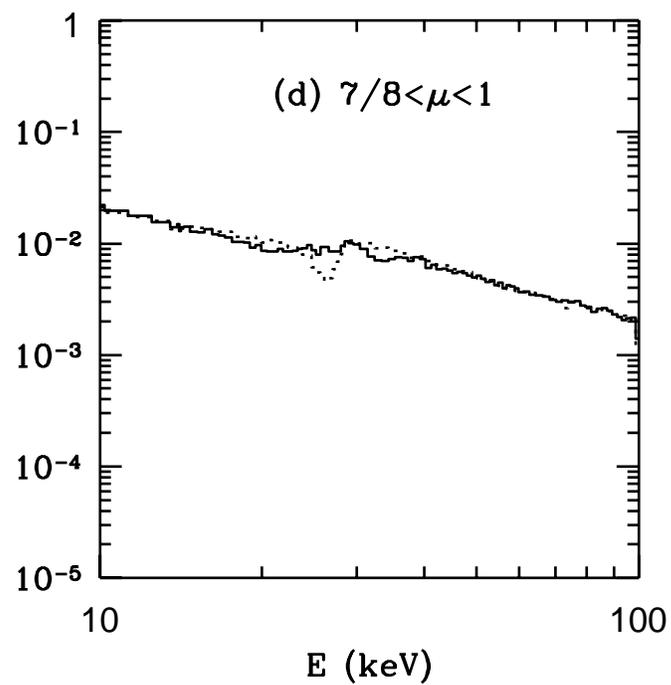
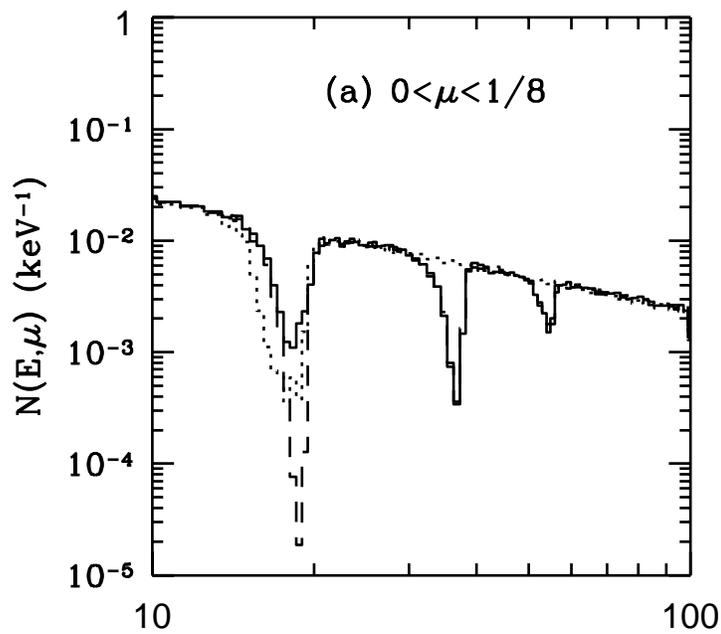
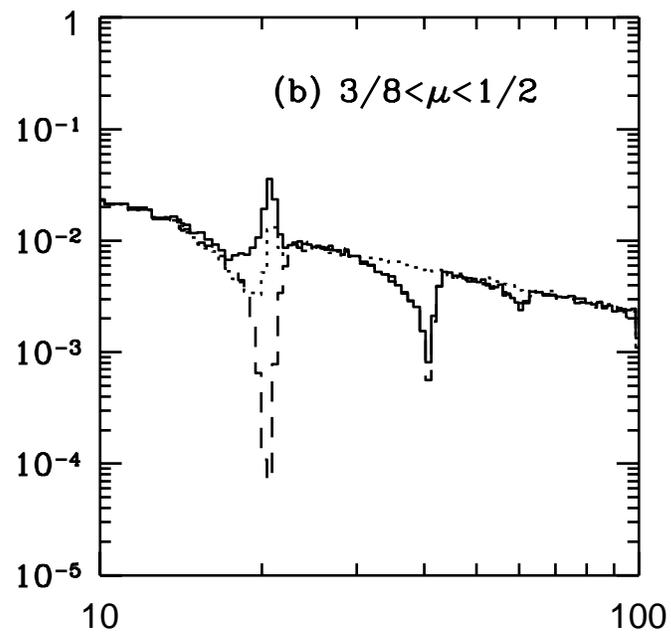
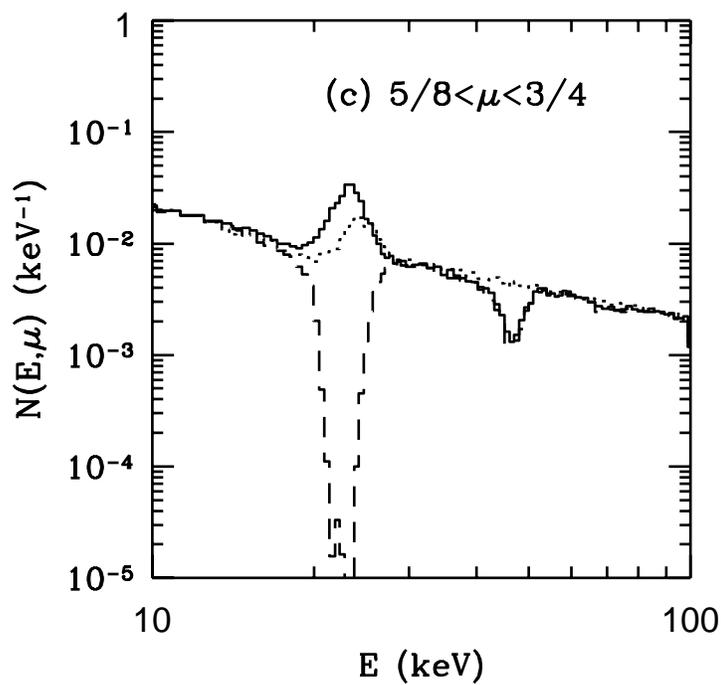
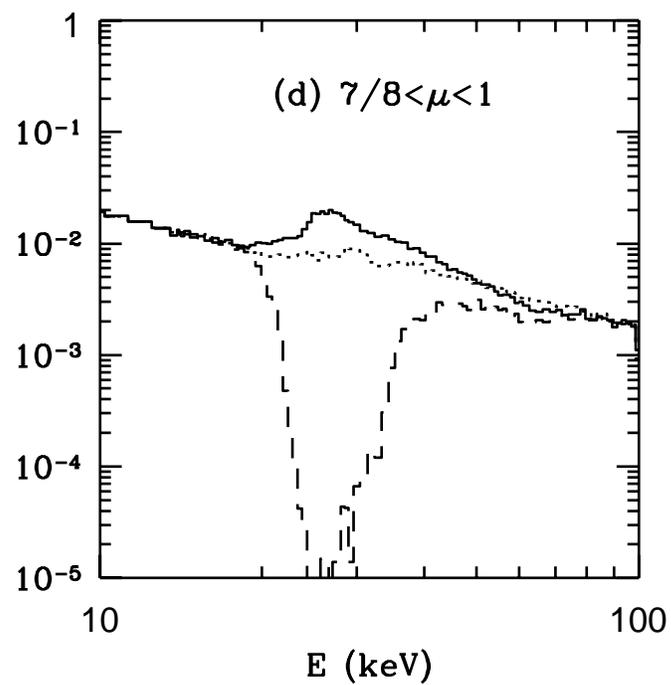
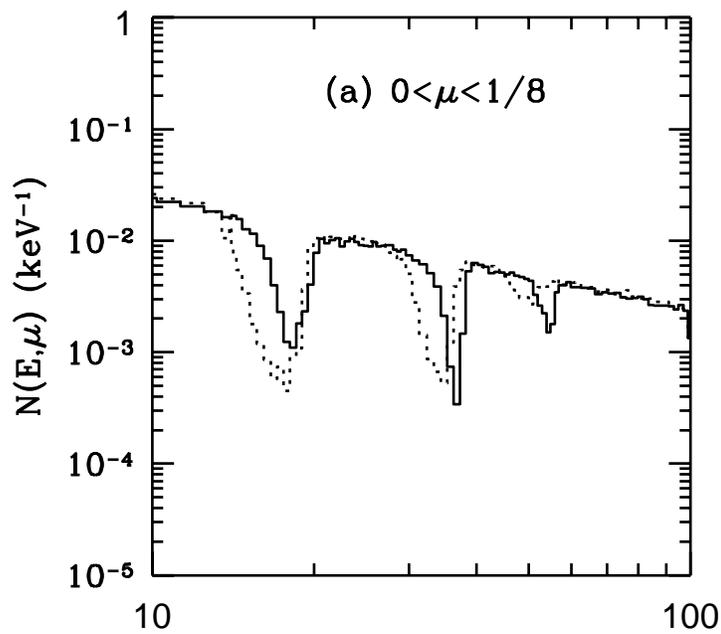
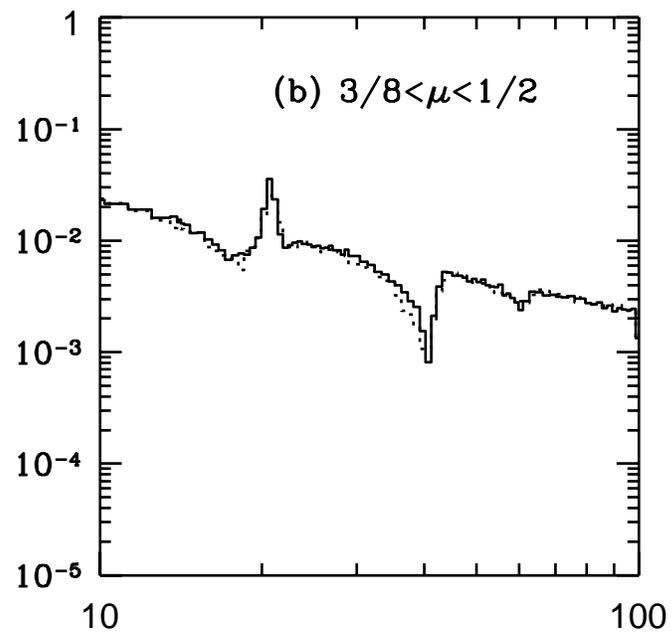
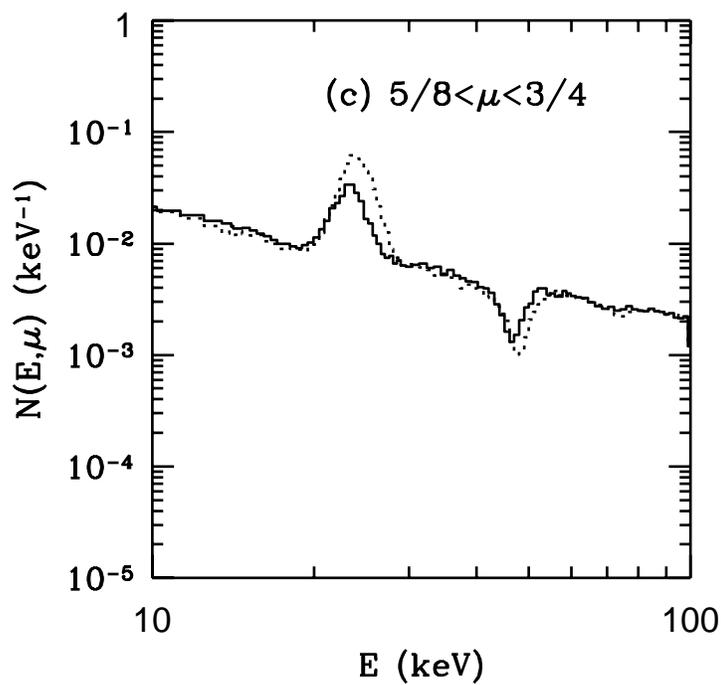
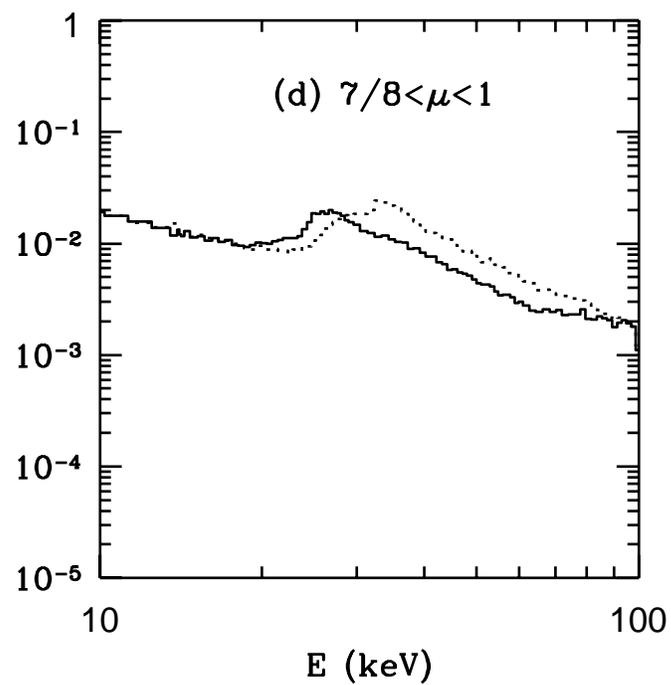
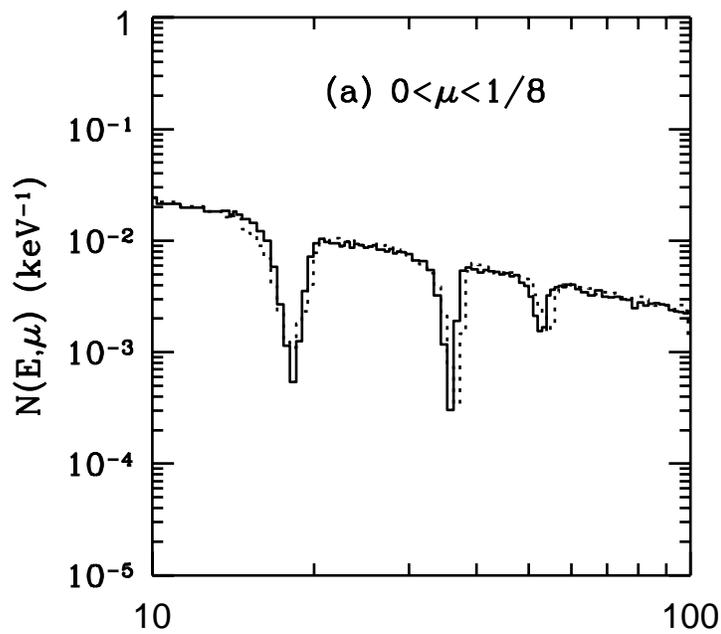
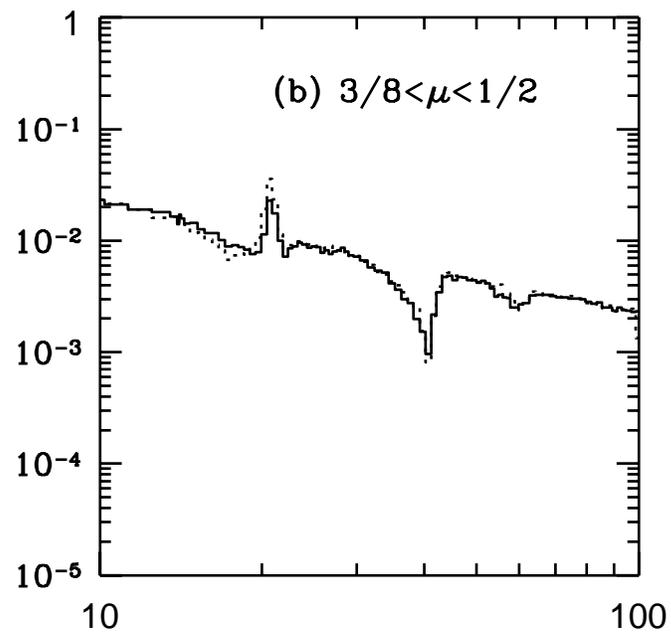
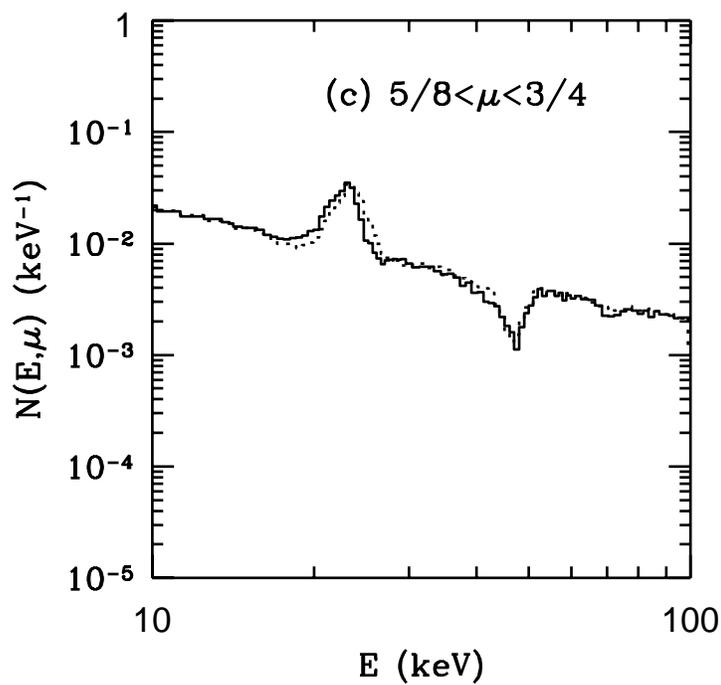
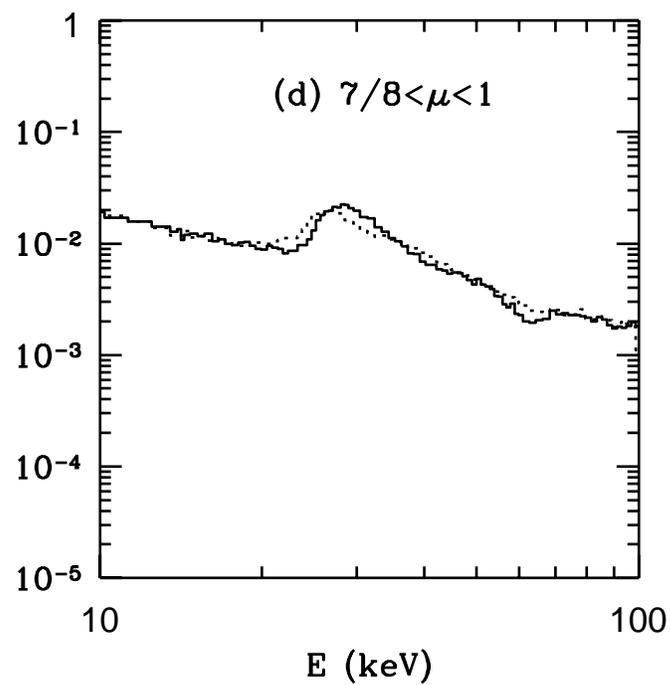
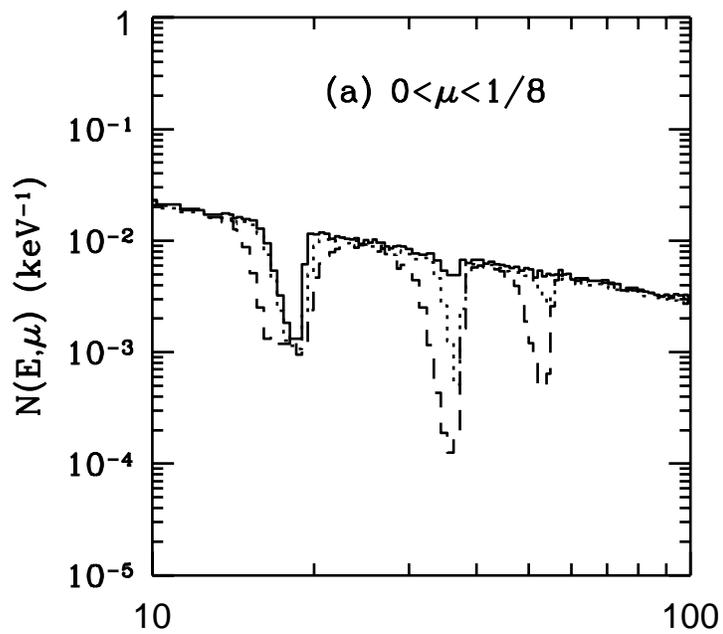
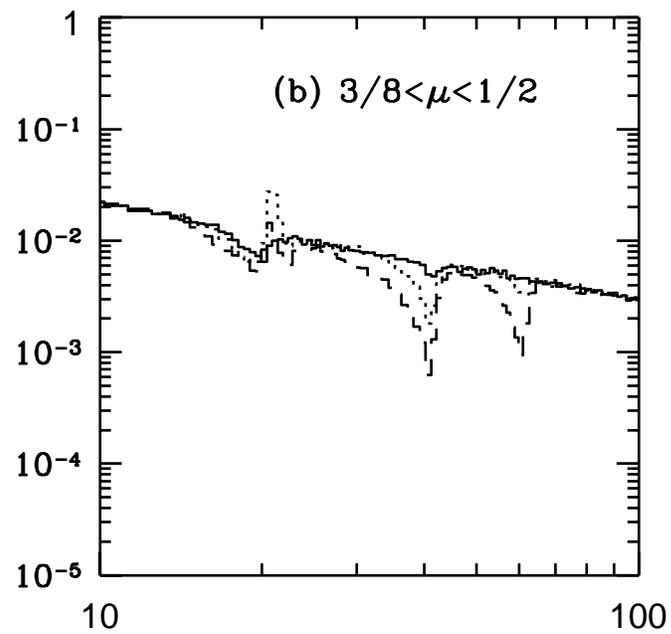
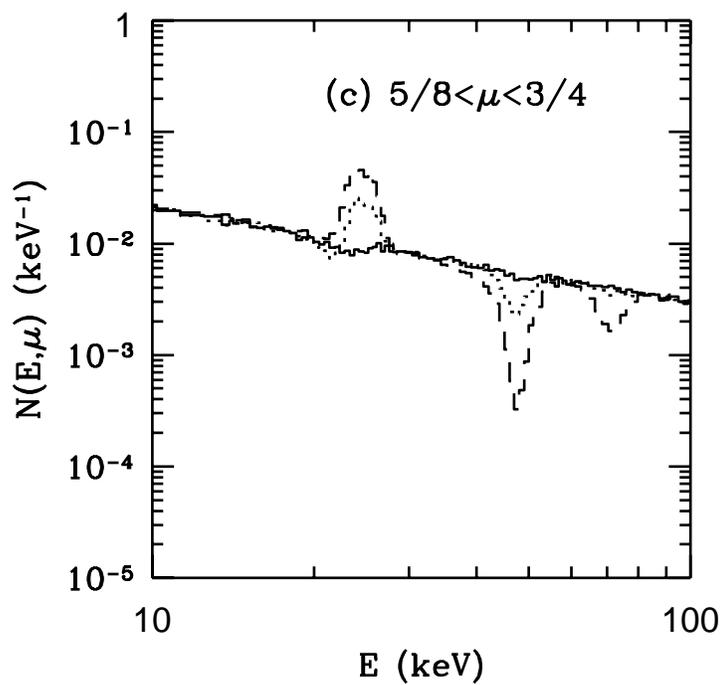
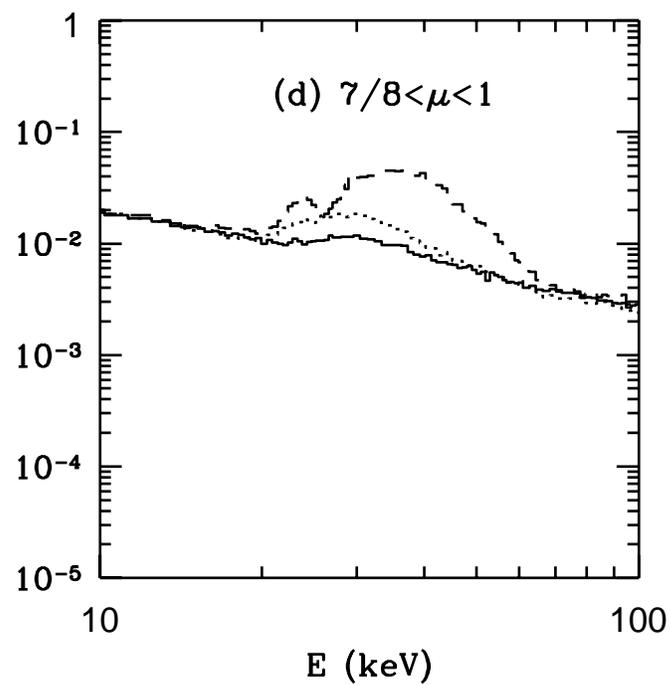
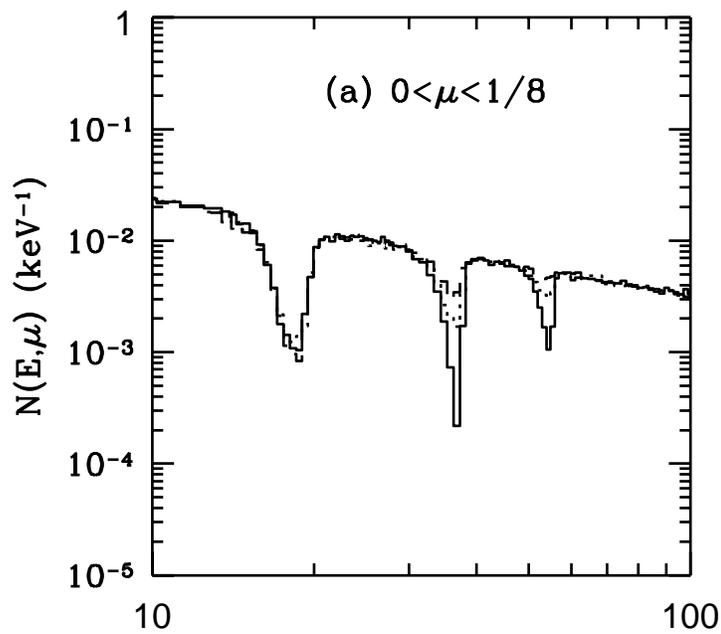
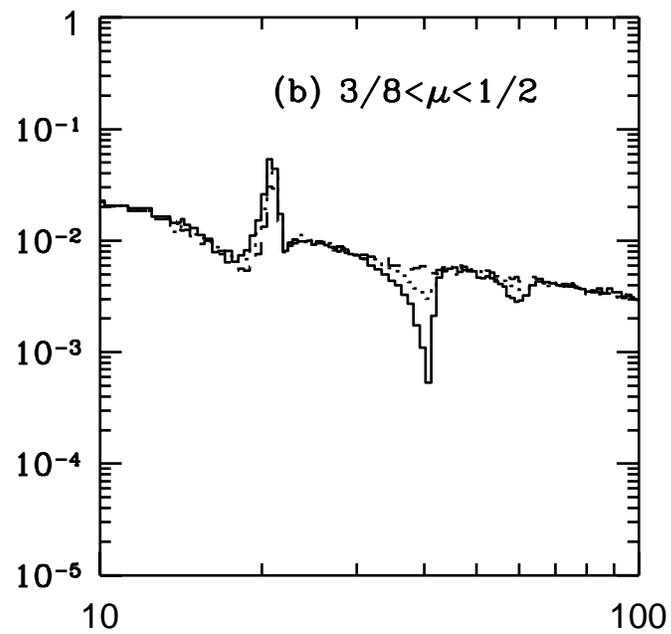
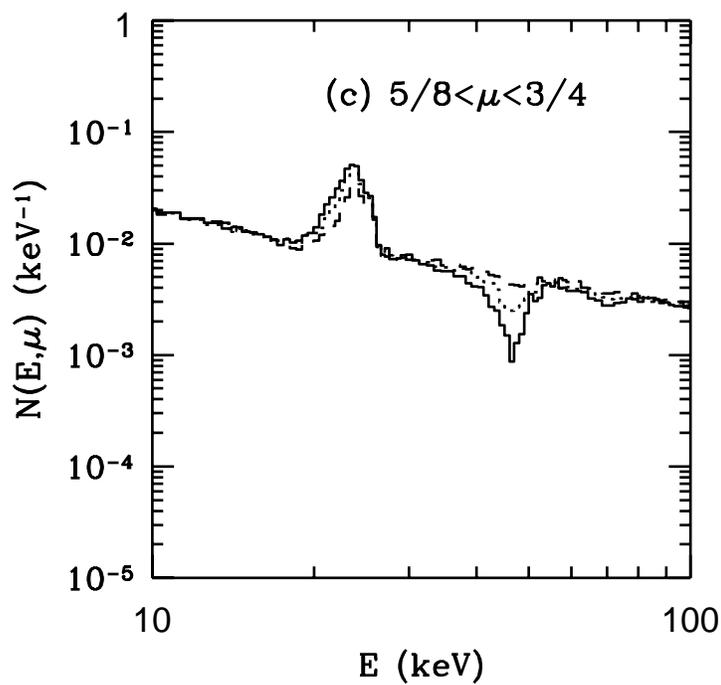
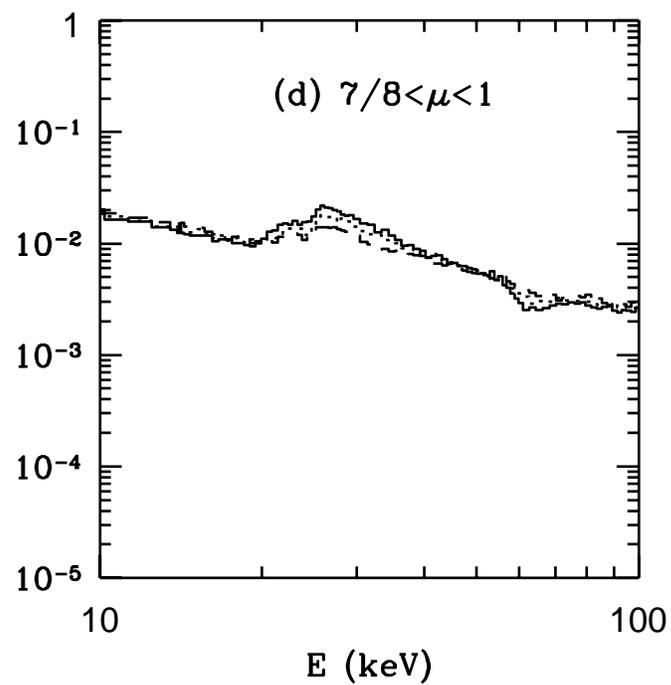
\end{document}